\documentclass{article}
\usepackage{upgreek}
\usepackage{latexsym,epsfig,bm,amssymb}
\usepackage{color}
\usepackage{mathptmx,amsthm,mathrsfs}
\input ot1ptm.fd \input ot1ztmcm.fd \input omlztmcm.fd
\input omsztmcm.fd  \input omxztmcm.fd \input ulasy.fd \input ursfs.fd

\newcommand{\notyet}[1]{}

\DeclareSymbolFont{AMSb}{U}{msb}{m}{n}
\DeclareSymbolFontAlphabet{\mathbb}{AMSb}

\newcommand\supp{\mathop{\rm supp}}
\newcommand\pa{\partial}
\newcommand\ov{\overline}
\newcommand\om{\omega}
\newcommand{\at}[1]{\vert\sb{\sb{#1}}}
\newcommand{\Spec}{\mathop{\rm Spec}}

\newcommand{\Bar}[1]{\mkern2mu\overline{\mkern-2mu#1\mkern-5mu}\mkern5mu}
\def\Re{{\rm Re\, }}
\def\Im{{\rm Im\,}}
\providecommand\C{{\mathbb C}}
\renewcommand\C{{\mathbb C}}
\newcommand{\R}{{\mathbb R}}
\newcommand{\N}{{\mathbb N}}

\newcommand{\const}{{\rm const}}
\newcommand\sothat{{\rm :}\ }
\newcommand\sgn{\mathop{\rm sgn}}

\font\thf cmssdc10 at 11pt

\theoremstyle{plain}
\newtheorem{theorem}{\thf Theorem}[section]

\newtheorem{lemma}[theorem]{\thf Lemma}
\newtheorem{corollary}[theorem]{\thf Corollary}
\newtheorem{proposition}[theorem]{\thf Proposition}

\theoremstyle{definition}
\newtheorem{definition}[theorem]{Definition}
\theoremstyle{remark}
\newtheorem{remark}[theorem]{\thf Remark}

\makeatletter\@addtoreset{equation}{section}
\makeatother

\begin{document}

\begin{center}
{\huge Global attractor for  1D Dirac field 
\bigskip\\
coupled to nonlinear oscillator}

\bigskip\bigskip

{\sc Elena Kopylova}
\footnote{Research supported by the Austrian Science Fund (FWF) under Grant No. P27492-N25 
and RFBR grant 18-01-00524.}
\\ 
{\it\small  Faculty of  Mathematics of Vienna University\\
\it\small Institute for Information Transmission Problems RAS
}
\smallskip\\
{\sc Alexander Komech}
\footnote{Research supported by the Austrian Science Fund (FWF) under Grant No. P28152-N35.
}
\\
{\it\small Faculty of  Mathematics of Vienna University\\
\it\small Institute for Information Transmission Problems RAS\\
\it\small Department Mechanics-Mathemcatics of Moscow State University
}

\end{center}

\date{}

\begin{abstract}
The long-time asymptotics is analyzed for all finite energy solutions to a model $\mathbf{U}(1)$-invariant
nonlinear Dirac equation in one dimension, coupled to a nonlinear oscillator:
{\it each finite energy solution} converges as $t\to\pm\infty$ to the set of all ``nonlinear eigenfunctions''
of the form $(\psi_1(x)e^{-i\omega_1 t},\psi_2(x)e^{-i\omega_2 t})$. The {\it global attraction} is caused by the nonlinear
energy transfer from lower harmonics to the continuous spectrum and subsequent dispersive radiation.

We justify this mechanism by the  strategy based on \emph{inflation of spectrum by the nonlinearity}.
We show that any {\it omega-limit trajectory} has the time-spectrum in the spectral gap $[-m,m]$ and
satisfies the original equation. This equation implies the key {\it spectral inclusion} for spectrum of the nonlinear term.
Then the application of the Titchmarsh convolution theorem reduces the spectrum of $j$-th component  of the omega-limit trajectory
to a single harmonic $\omega_j\in[-m,m]$, $j=1,2$.
\end{abstract}
\section{Introduction}
\label{sect-introduction}
We prove  a global attraction for a model 1D nonlinear Dirac equation
\begin{equation}\label{KG-0}
i\dot\psi(x,t)=D_m\psi(x,t)-D_m^{-1}\delta(x)F(\psi(0,t)),\quad x\in\R.
\end{equation}
Here $D_m$ is the Dirac operator  $D_m:=\alpha \partial_x+m\beta $,
where  $m>0$,  and
$$
\alpha=\left(\begin{array}{cc} 0 & 1\\-1& 0\end{array}\right), \quad\beta=\left(\begin{array}{cc} 1 & 0 \\0& -1\end{array}\right),
$$ 
 $\psi(x,t)=(\psi_1(x,t),\psi_2(x,t))$ is a continuous $C^2$-valued wave function, and $F(\zeta)=(F_1(\zeta_1), F_2(\zeta_2))$, $\zeta=(\zeta_1,\zeta_2)\in \C^2$, 
 is a  nonlinear function. The dots stand for the derivatives in $t$. All derivatives and the equation are understood in the sense of distributions.
Equation (\ref{KG-0}) describes the linear Dirac equation coupled to the nonlinear oscillator with the interaction concentrated in one point.
We assume that equation (\ref{KG-0}) is $\mathbf{U}(1)$-invariant; that is,
\begin{equation}\label{inv-f}
F(e^{i\theta}\zeta)=e^{i\theta}F(\zeta),\qquad\zeta\in\C^2,\quad\theta\in\R.
\end{equation}
This condition  leads to the existence of  {\it two-frequency} solitary wave solutions of type
\begin{equation}\label{soliton}
\psi(x,t)=\phi_{\omega_1}(x)e^{-i\omega_1 t}+\phi_{\omega_2}(x)e^{-i\omega_2 t},\quad (\omega_1,\omega_2)\in\R^2.
\end{equation}
We prove that indeed they form the global attractor for all finite energy solutions to (\ref{KG-0}).
Namely, our main result is the following long-time asymptotics: In the case when polynomials  $F_j$  are {\it strictly nonlinear},  
any  solution with initial data from $H^1(\R)\otimes C^2$ converges to the set $\mathcal{S}$ of all solitary waves:
\begin{equation}\label{attraction}
\psi(\cdot,t)\longrightarrow\mathcal{S},
\qquad t\to\pm\infty,
\end{equation}
where the convergence holds in local  $H^1$ - seminorms.
\medskip\\
The asymptotics of type (\ref{attraction}) was discovered first for linear wave and Klein-Gordon equations with external potential
in the scattering theory \cite{Str68,MS72,Kla82,Hor91}.
In this case, the attractor $\mathcal{S}$ consists of the zero solution only,
and the asymptotics means well-known \emph{local energy decay}.

The  attraction to the set of all \emph{static} stationary states with $\omega=0$ was established in
\cite{Kom95}--\cite{KV96} for  a number nonlinear wave problems.  

First results on the attraction  to the set of all stationary orbits  for nonlinear $\mathbf{U}(1)$-invariant Schr\"o\-din\-ger equations
were obtained in the context of asymptotic stability. This establishes asymptotics of type (\ref{attraction}) but only for solutions 
with initial date close to some stationary orbit, proving the existence of a \emph{local attractor}.
This was first done in \cite{BP95,SW90,SW92}, and then developed in \cite{ANO, BS03,BKKS, Cuc01, KKS12} and other papers.

The  global attraction of type (\ref{attraction}) to the solitary waves was established
i) in \cite{KK07, KK2010a} for 1D Klein-Gordon equations coupled to  nonlinear oscillators;
ii) in \cite{KK2009, KK2010} for  nD Klein-Gordon and Dirac equations with mean field  interaction;
iii)  in \cite{K16, K17}  for  3D wave and Klein-Gordon   equations  with concentrated  nonlinearity.

The \emph{global attraction} (\ref{attraction})  for Dirac equation
 with concentrated  nonlinearity
 was not considered previously as well as the attraction to  solitary waves with  two frequencies. 
 
 \smallskip

Let us comment on our methods.
First we prove the {\it omega-limit compactness}. This means that for each sequence $s_j\to\infty$
the solutions $\psi(x,t+s_j)$ contain an infinite  subsequence which converges in energy seminorms for $|x|<R$
 and $|t|<T$ for any $R,T>0$. Any limit function is called as the {\it omega-limiting trajectory} $\gamma(x,t)$. To prove the global convergence 
 (\ref{attraction}) is suffices to show that any omega-limiting trajectory lies on  $\mathcal{S}$.
 
 The proof relies on the study of the Fourier transform in time $\tilde\psi(x,\om)$ and of its support which is the 
 time-spectrum.
 The key role is played by the 
 absolute continuity of the spectral densities  $\tilde\psi(x,\om)$ outside the spectral gap $[-m,m]$ for each $x\in\R$.
 The absolute continuity is a nonlinear version of Kato's theorem on the absence
of the embedded eigenvalues and provides the dispersion decay for the high energy
component. Any omega-limit trajectory  is the solution to (\ref{KG-0}).   

This  absolute continuity provides that the time-spectrum  of $\tilde\gamma(x,\om)=(\tilde\gamma_1(x,\om),\tilde\gamma_2(x,\om))$ 
is contained in the spectral gap $[-m,m]$ for each $x\in\R$.
Finally, we apply the Titchmarsh convolution theorem (see  \cite[Theorem 4.3.3]{Hor90})
to conclude that each components $\tilde\gamma_j(x,\om)$ of omega-limit trajectory is a singleton, 
i.e. it's time-spectrum  consists of a single frequency.
The Titchmarsh theorem controls the inflation of spectrum by the nonlinearity. 
Physically, these arguments justify the following binary mechanism of the
energy radiation, which is responsible for the attraction to the solitary waves: 
(i)  nonlinear energy transfer from the lower to higher harmonics, 
and (ii)  subsequent dispersion decay caused by the energy radiation to infinity.

The general scheme of the proof bring to mind the approach of \cite{KK07,KK2009}.
Nevertheless, the Dirac equation  with nonlinear point interaction
 requires new ideas due to a more singular character. In particular, the equation with 
 nonlinearity $F(\psi(0,t))\delta(x)$ seems not well posed.
 Indeed, the solution of the free 1D Dirac equation with the source 
 $f(t)\delta(x)$ does not belong to  $H^1(\R)$, so  $\psi(0,t)$ is not well defined.

 We found the novel model of Dirac equation with
non-standard Hamilton structure, which provides good a priori estimates  needed for
the global well-posedness. As a consequence, the formulation of the problem and the techniques used 
are not a straightforward generalization of the one-dimensional result \cite{KK07}.

The plan of the paper is as follows.
In Section~\ref{sect-results} we state the main assumptions and results.
In Section~\ref{sect-splitting} we eliminate a dispersive component
of the solution and construct spectral representation for the remaining part. 
In Section~\ref{abs_con} we prove absolute continuity of high frequency spectrum of the remaining part.
In Section~\ref{decom-2} we exclude the second dispersive component corresponding to the high frequencies.
In Section~\ref{sect-bound} we establish compactness for the remaining  component with the bounded spectrum. 
In Section~\ref{sect-spectral} we state the spectral properties of all omega-limit trajectories
and apply the Titchmarsh Convolution Theorem.
In Appendices we establish the global well-posedness for equation (\ref{KG-0}) and prove global attraction (\ref{attraction})
in the case on  linear $F(\psi)$.
\section{Main results}
\label{sect-results}
\subsection*{Model}
We consider the Cauchy problem for the Dirac equation coupled to a nonlinear oscillator: 
\begin{equation}\label{D}
\left\{\begin{array}{l}
i\dot\psi(x,t)=D_m\psi(x,t)-D_m^{-1}\delta(x)F(\psi(0,t)), \quad t\in\R
\\
\psi\at{t=0}=\psi\sb 0(x),
\end{array}\right|\quad x\in\R.
\end{equation}
We will assume that the nonlinearity $F=(F_1,F_2)$ admits a real-valued potential:
\begin{equation}\label{P}
F_j(\zeta)=-\pa_{\ov\zeta\!_j} U(\zeta),\quad\zeta_j\in\C, \quad j=1,2, \qquad U\in C^2(\C^2).
\end{equation}
Then equation (\ref{D}) formally can be written as a Hamiltonian system,
\[
\dot\psi(t)=J\,D\mathcal{H}(\psi),\qquad J=-iD_m^{-1},
\]
where $D\mathcal{H}$ is the variational derivative of the Hamilton functional
\begin{equation}\label{hamiltonian}
\mathcal{H}(\psi)=\frac 12\langle \psi, (-\partial_x^2+m^2)\psi\rangle+U(\psi(0)).
\end{equation}

\subsection*{Global well-posedness}
To have a priori estimates available for the proof of the global well-posedness, we assume that
\begin{equation}\label{bound-below}
U(\zeta)\ge A-B|\zeta|^2, \qquad \zeta\in\C^2,\qquad  A\in\R,\qquad 0\le B<m,
\end{equation}
We will write $L^2$ and $H^1$ instead of $L^2(\R)\otimes C^2$ and instead of $H^1(\R)\otimes C^2$, respectively.
\begin{theorem}\label{theorem-well-posedness}
Let  conditions (\ref{P}) and (\ref{bound-below}) hold. Then:
\begin{enumerate}
\item
For every $\psi_0\in H^1$  the Cauchy problem (\ref{D}) has a unique solution $\psi(t)\in C(\R,H^1)\cap C(\R,L^2)$.
\item
The map $W(t):\;\psi_0\mapsto \psi(t)$ is continuous in $H^1$  for each $t\in\R$.
\item
The energy is conserved: 
\begin{equation}\label{ec}
\mathcal{H}(\psi(t))=\const,\quad t\in\R.
\end{equation}
\item
The following \emph{a priori} bound holds: 
\begin{equation}\label{eb}
\Vert \psi(t)\Vert_{H^1}\le C(\psi_0),\qquad t\in\R.
\end{equation}
\end{enumerate}
\end{theorem}
We prove this theorem in Appendix~\ref{sect-existence}.
\subsection*{Solitary waves and the main theorem}
We assume that the nonlinearity is polynomial.  More precisely,
\begin{equation}\label{inv-u}
U(\zeta)=U_1(\zeta_1)+U_2(\zeta_2),\qquad\zeta_j\in\C,
\end{equation}
where
\begin{equation}\label{inv-u1}
U_j(\zeta_j)=\sum\limits_{n=0}^{N_j}u_{n,j}|\zeta_j|^{2n},\quad u_{n,j}\in\R, \quad u_{N_j,j}>0, \quad N_j\ge 2,\quad j=1,2.
\end{equation}
This assumption guarantees the bound (\ref{bound-below}) and 
it  is crucial in our argument: it  allow to apply the Titchmarsh convolution theorem. 
Equality (\ref{inv-u1}) implies that
\begin{equation}\label{def-a}
F_j(\zeta_j)=-\partial_{\overline\zeta_j} U_j(\zeta_j)=a_j(|\zeta_j|^2)\zeta_j,\quad j=1,2,
\end{equation}
\begin{equation}\label{aj-def}
a_j(|\zeta_j|^2):=-\sum\limits_{n=1}^{N_j}2nu_{n,j}|\zeta_j|^{2n-2}.
\end{equation}
\begin{definition}\label{soldef}
(i) The solitary wave solution of equation  (\ref{KG-0}) are solutions of the form
\begin{equation}\label{solv}
\psi(x,t)=\phi_{\omega_1}(x)e^{-i\omega_1 t}+\phi_{\omega_2}(x)e^{-i\omega_2 t},\quad (\omega_1,\omega_2)\in\R^2,\quad \phi_{\omega_k}\in H^1,~~k=1,2.
\end{equation}
(ii) The solitary manifold is the set:
${\cal S}=\left\{\phi_{\omega_1}+\phi_{\omega_2}\sothat~~(\omega_1,\omega_2)\in\R^2\right\}$.
\end{definition}
Note that   for any $(\omega_1,\omega_2)\in\R^2$ there is a zero solitary wave with $\phi_{\omega_1}=\phi_{\omega_2}\equiv 0$,  since $F(0)=0$.
From (\ref{inv-u}) it follows  that the set ${\cal S}$ is invariant under multiplication by 
$e\sp{i\theta}$, $\theta\in\R$.

Denote $\varkappa_j=\varkappa(\omega_j)=\sqrt{m^2-\omega_j^2}> 0$ for $\omega_j\in (-m,m)$.
\begin{proposition}\label{sol-ex} (Existence of solitary waves).
Assume that $F(\zeta)$ satisfies (\ref{def-a}). Then nonzero solitary waves may exist only 
for $\omega_j\in (-m,m)$. The amplitudes of solitary waves are given by 
\begin{equation}\label{psisol}
\phi_{\omega_1}(x)=C_1\!\left(\!\!\begin{array}{cc} e^{-\varkappa_1|x|}+\frac{me^{-\varkappa_1|x|}-\varkappa_1e^{-m|x|}}{\omega_1}\\ 
\varkappa_1\sgn x\frac{e^{-\varkappa_1|x|}-e^{-m|x|}}{\omega_1}\end{array}\!\!\right)\!,~~
\phi_{\omega_2}(x)=C_2\!\left(\!\!\begin{array}{cc} -\varkappa_2\sgn x\frac{e^{-\varkappa_2|x|}-e^{-m|x|}}{\omega_2}\\
e^{-\varkappa_2|x|}-\frac{me^{-\varkappa_2|x|}-\varkappa_2e^{-m|x|}}{\omega_2} \end{array}\!\!\right)\!,~~\omega_j\in (-m,m),~~ j=1,2,
\end{equation}
where $C_j$  are  solutions to 
\begin{equation}\label {qsol}
2C_j\varkappa_j=F_j\Big(C_j\Big[1+(-1)^{j+1}\frac{m-\varkappa(\omega_j)}{\omega_j}\Big]\Big),\quad j=1,2.
\end{equation}
\end{proposition}
\begin{corollary}
Substituting (\ref{psisol}) into (\ref{solv}) we obtain the following representation for solitary wave solutions
\begin{equation}\label{psiOmsol}
\left\{ \begin{array}{cc}
\psi_1(x,t)=C_1\Big(e^{-\varkappa_1|x|}+\frac{m e^{-\varkappa_1|x|}-\varkappa_1e^{-m|x|}}{\omega_1}\Big)e^{-i\omega_1 t}
-C_2\varkappa_2\sgn x\frac{e^{-\varkappa_2|x|}-e^{-m|x|}}{\omega_2}e^{-i\omega_2 t}\\\\
\psi_2(x,t)=C_1\varkappa_1\sgn x{\omega_1}\frac{e^{-\varkappa_1|x|}-e^{-m|x|}}{\omega_1}e^{-i\omega_1 t}
+C_2\Big(e^{-\varkappa_2|x|}-\frac{m e^{-\varkappa_2|x|}-\varkappa_2 e^{-m|x|}}{\omega_2}\Big)e^{-i\omega_2 t}
\end{array}\right|
\end{equation}
\end{corollary}
{\bf Proof of Proposition \ref{sol-ex}}
We look for solution  $\psi(x,t)$  to (\ref{KG-0}) in the form (\ref{solv}). Consider the function
\begin{equation}\label{xi-def}
\chi(x,t):=\psi(x,t)-iD_m^{-1}\dot\psi(x,t)=\chi_{\omega_1}(x)e^{-i\omega_1 t}+\chi_{\omega_2}(x)e^{-i\omega_2 t},
\end{equation}
where
$$
\chi_{\omega_k}=\phi_{\omega_k}-\omega_k D_m^{-1}\phi_{\omega_k}=D_m^{-1}(D_m-\omega_k)\phi_{\omega_k},\qquad k=1,2.
$$
Hence, 
\begin{equation}\label{psi-om-rep}
\phi_{\omega_k}=D_m (D_m-\omega_k)^{-1}\chi_{\omega_k}=D_m (D_m+\omega_k)(D_m^2-\omega_k^2)^{-1}\chi_{\omega_k}
=\chi_{\omega_k}+(\omega_k^2+\omega_k D_m)(D_m^2-\omega_k^2)^{-1}\chi_{\omega_k}.
\end{equation}
Equation (\ref{KG-0}) implies, that
$$
D_m \chi(x,t)=D_m\psi(x,t)-i\dot\psi(x,t)=D_m^{-1}F(\psi(0,t))\delta(x).
$$
Applying   the operator $D_m$, we obtain by (\ref{xi-def})
$$
e^{-i\omega_1 t}D_m^2 \chi_{\omega_1}(x)+e^{-i\omega_2 t}D_m^2 \chi_{\omega_2}(x)=F(\psi(0,t))\delta(x), \quad
D_m^2=-\pa^2_x+m^2.
$$
Therefore, in the case $\om_1\ne\om_2$,
\begin{equation}\label{chi-om-sol}
\chi_{\omega_k,j}(x)=C_{kj}\frac{e^{-m|x|}}{2m},\quad k,j=1,2,
\end{equation}
where $C_{kj}$ are  solutions to
\begin{eqnarray}\label{Ckj-sol}
e^{-i\omega_1 t}C_{1j}+e^{-i\omega_2 t}C_{2j}
=F_j(\phi_{\omega_1,j}(0)e^{-i\omega_1 t}+\phi_{\omega_2,j}(0)e^{-i\omega_2 t}),\quad j=1,2.
\end {eqnarray} 
We can also assume this formulas
in the case $\om_1=\om_2$ setting $\chi_{\omega_2}=0$. We will return to  equation (\ref{Ckj-sol}) later.
First we derive the explicit formulas for $\phi_{\omega_k}(x)$, using (\ref{psi-om-rep}) and (\ref{chi-om-sol}) only. 
Applying \cite [Formula  1.2.(11)]{E}, we get
\begin{eqnarray}\nonumber
(D_m^2-\omega_k^2)^{-1}\frac{C_{kj}e^{-m|x|}}{2m}&=&\frac{C_{kj}}{2\pi}\int _{\R}\frac{e^{-i\xi x} d\xi}{(\xi^2+m^2)(\xi^2+m^2-\omega_k^2)}
=\frac{C_{kj}}{\omega_k^2\pi}\int _0^{\infty}\Big(\frac{\cos \xi x}{\xi^2+m^2-\omega_k^2}-\frac{\cos \xi x}{\xi^2+m^2}\Big) d\xi\\
\nonumber
&=&\frac{C_{kj}}{2\omega_k^2}\Big(\frac{e^{-\varkappa_k|x|}}{\varkappa_k}-\frac{e^{-m|x|}}{m}\Big),\quad\varkappa_k=\varkappa(\omega_k). 
\end{eqnarray}
Substituting this into (\ref{psi-om-rep}), we obtain
\begin{equation}\label{phix}
\left(\begin{array}{cc} \phi_{\omega_k,1}(x)\\ \phi_{\omega_k,2}(x)\end{array}\right)=
\left(\begin{array}{cc} C_{k1}\\ C_{k2}\end{array}\right)\frac{e^{-\varkappa_k|x|}}{2\varkappa_k}
+\left(\begin{array}{cc} C_{k1}\\ -C_{k2}\end{array}\right)\Big(\frac{me^{-\varkappa_k|x|}-\varkappa_ke^{-m|x|}}{2\omega_k\varkappa_k}\Big)
+\left(\begin{array}{cc} -C_{k2}\\ C_{k1}\end{array}\right)\sgn x\frac{e^{-\varkappa_k|x|}-e^{-m|x|}}{2\omega_k}.
\end{equation}
Hence,
\begin{equation}\label{phi0}
\phi_{\omega_k,j}(0)=\frac{C_{kj}}{2\varkappa_k}\big(1+(-1)^{j+1}\frac{m-\varkappa_k}{\omega_k}\big),\quad k,j=1,2.
\end{equation}
Now we turn  to the study of the equation (\ref{Ckj-sol}).
First, consider the case when $\omega_1=\omega_2=\omega$. We set $C_{1j}=C_j$,  $C_{2j}=0$, $j=1,2$,
and equation (\ref{Ckj-sol}) becomes
$$
e^{-i\omega t}C_{j}=F_j(\phi_{\omega,j}(0)e^{-i\omega t})
=a_j (|\phi_{\omega,j}(0)e^{-i\omega t}|)\phi_{\omega,j}(0)e^{-i\omega t},\quad j=1,2.
$$
by (\ref{def-a}). Using (\ref{phi0}), we get after cancelation of exponential 
\begin{eqnarray}\label{Ckj-sol-lin}
2C'_{j}\varkappa=F_j(\phi_{\omega,j}(0))=F_j(C'_{j}\big(1+(-1)^{j+1}\frac{m-\varkappa}{\omega}\big)),\quad\varkappa=\sqrt{m^2-\omega^2},\quad j=1,2.
\end {eqnarray}
Here we denote $C_j'=C_j/(2\varkappa)$. Finally, in the case $\omega_1=\omega_2=\omega$, equation (\ref{solv}) reads
$$
\psi(x,t)=\phi_{\omega}(x,t)e^{-i\omega t}, 
$$
where, in accordance with  (\ref{phix}),
\begin{equation}\label{phix0}
\left\{ \begin{array}{cc}
\phi_{\omega,1}(x)=C'_{1}\Big( e^{-\varkappa|x|}+\frac{me^{-\varkappa|x|}-\varkappa e^{-m|x|}}{\omega}\Big)
-C'_{2}\varkappa\sgn x\frac{e^{-\varkappa|x|}-e^{-m|x|}}{2\omega_k}\\\\
\phi_{\omega,2}(x)=C'_{2}\Big (e^{-\varkappa|x|}-\frac{me^{-\varkappa|x|}-\varkappa e^{-m|x|}}{\omega}\Big)
+C'_1\varkappa\sgn x\frac{e^{-\varkappa|x|}-e^{-m|x|}}{2\omega_k}
\end{array}\right|
\end{equation}
Now consider the case when $\omega_1\not =\omega_2$. Taking into account (\ref{def-a}), we rewrite (\ref{Ckj-sol}) as
\begin{eqnarray}\label{Ckj-sol1}
e^{-i\omega_1 t}C_{1j}+e^{-i\omega_2 t}C_{2j}
=a_j (|\phi_{\omega_1,j}(0)e^{-i\omega_1 t}+\phi_{\omega_2,j}(0)e^{-i\omega_2 t}|^2)
(\phi_{\omega_1,j}(0)e^{-i\omega_1 t}+\phi_{\omega_2,j}(0)e^{-i\omega_2 t}),\quad j=1,2.
\end {eqnarray}
\begin{lemma}\label{rem-lem}
Let $\omega_1\ne \omega_2$. Then  for solutions to
(\ref{Ckj-sol}) we have  either $\phi_{\omega_1,j}(0)=0$ or $\phi_{\omega_2,j}(0)=0$ for each $j=1,2$.
\end{lemma}
\begin{proof}
It suffices to consider the case $j=1$ and $\om_1<\om_2$ only. 
Denote $q_1:=\phi_{\omega_1,1}(0)$,  $q_2:=\phi_{\omega_2,1}(0)$. 
We should prove that either $q_1=0$ or $q_2=0$. Assume, to the  contrary, that $q_1\ne 0$ and $q_2\ne 0$.
Then 
$$
|\phi_{\omega_1,1}(0)e^{-i\omega_1 t}+\phi_{\omega_2,1}(0)e^{-i\omega_2 t}|^2=
|q_1|^2+|q_2|^2+{q}_1\ov q_2 e^{i\delta t}+\ov q_1 {q}_2e^{-i\delta t},\quad\delta:=\omega_2-\omega_1>0,
$$
where ${q}_1\ov q_2\ne 0$ and $\ov {q}_1 q_2\ne 0$.
Hence,  (\ref{aj-def}) implies
$$
a_1 (|\phi_{\omega_1,1}(0)e^{-i\omega_1 t}+\phi_{\omega_2,1}(0)e^{-i\omega_2 t}|^2)
=be^{i(N_j-1)\delta t}+\ov be^{-i(N_j-1)\delta t}+\sum\limits_{|n|\le N_1-2}c_ne^{in\delta t}, 
$$
where $b\ne 0$
since $a_1$ is a polynomial of degree $N_1-1\ge 1$ due to (\ref{inv-u1}) and (\ref{aj-def}).
Now the right hand side of (\ref{Ckj-sol}) 
contains the terms $e^{-i[\om_1t-(N_j-1)\delta] t}$ and $e^{-i[\om_2t+(N_j-1)\delta] t}$
with nonzero coefficients, which are absent on the left hand side.
This contradiction proves the lemma.
\end{proof}
This lemma  and  (\ref{phi0}) imply
\begin{corollary}
Let $\omega_1\ne \omega_2$. Then  for  solutions to (\ref{Ckj-sol}) we have  either $C_{1,j}=0$ or $C_{2,j}=0$ for each $j=1,2$.
\end{corollary}
Note that the  cases  $C_{21}=C_{22}=0$ and  $C_{11}=C_{12}=0$ are exactly the case when $\omega_1=\omega_2$.\\
Suppose now, that $C_{12}=C_{21}=0$. Then  (\ref{solv}) and (\ref{phix}) imply
 \begin{equation}\label{eq2}
\left(\begin{array}{cc} \psi_{1}(x,t)\\ \psi_{2}(x,t)\end{array}\right)
=\frac{C_{1}}{2\varkappa_1}\left(\begin{array}{cc} e^{-\varkappa_1|x|}+\frac{me^{-\varkappa_1|x|}-\varkappa_1e^{-m|x|}}{\omega_1}\\ 
\varkappa_1\sgn x\frac{e^{-\varkappa_1|x|}-e^{-m|x|}}{\omega_1}\end{array}\right)e^{-i\omega_1 t}
+\frac{C_{2}}{2\varkappa_2}\left(\begin{array}{cc} -\varkappa_2\sgn x\frac{e^{-\varkappa_2|x|}-e^{-m|x|}}{\omega_2}\\
e^{-\varkappa_2|x|}-\frac{me^{-\varkappa_2|x|}-\varkappa_2e^{-m|x|}}{\omega_2} \end{array}\right)e^{-i\omega_2 t},
\end{equation}
with $C_1=C_{11}$, $C_2=C_{22}$. Taking into account (\ref{phi0}), we obtain  equations for $C_j$:
 \begin{equation}\label{eq3}
 C_{j}=F_j(\frac{C_{j}}{2\varkappa_j}[1+(-1)^{j+1}\frac{m-\varkappa_j}{\omega_j}]),\quad j=1,2. 
\end{equation}
Equations (\ref{eq2})  and  (\ref{eq2}) will coincide with equations (\ref{psiOmsol})  and (\ref{qsol}) 
after   the replacement $C_{j}$ by $2C_j\varkappa_j$.\\
It is easy to check that in the case $C_{11}=C_{22}=0$,  we obtain the same formulas, interchanging $\omega_1$ and $\omega_2$.\\
Proposition is completely proved.
\begin{remark}
({\it i}) Equation (\ref{qsol}) has generally discrete set of solutions $C_j$, while $\omega_j$ belongs generally to an open set.
\\
({\it i}) In the linear case, when $F_j(\psi_k)=a_j\psi_j$ with $a_j\in\R$, the situation is contrary :  we see from (\ref{qsol}) that 
$$
2\varkappa_j=a_j\Big(1+(-1)^{j+1}\frac{m-\varkappa_j}{\omega_j}\Big),\quad \varkappa_j=\sqrt{m^2-\omega_j^2},
$$
i.e., $\omega_j$ generally belongs to a discrete set, while $C_j\in\C$ is arbitrary. 
\end{remark}
Our main result is the following theorem.
\begin{theorem}[Main Theorem]
\label{main-theorem}
Let the nonlinearity $F(\psi)$ satisfy Assumption~\ref{assumption-a}. Then for any $\psi_0\in H^1$
the solution $\psi(t)\in C(\R, H^1)$ to the Cauchy problem {\rm (\ref{D})} with $\psi(0)=\psi_0$
converges to ${\cal S}$ in the space $H^1_{loc}(\R)\otimes C^2$:
\begin{equation}\label{cal-A}
\psi(t)\to {\cal S},\quad t\to \pm\infty.
\end{equation}
\end{theorem}
\section{Splitting of solutions}
\label{sect-splitting}
It suffices to prove Theorem~\ref{main-theorem} for $t\to+\infty$; 
We will only consider the solution $\psi(x,t)$ restricted to $t\ge 0$ and split it   as
\[
\psi(x,t)=\phi(x,t)+\psi_{S}(x,t),\quad t\ge 0.
\]
Here $\phi(x,t)$   is a solution to the  Cauchy problem for the free Dirac equation
\begin{equation}\label{D-1}
i\dot\phi(x,t)=D_m\phi(x,t),
\qquad \phi\at{t=0}=\psi_{0},
\end{equation}
and $\psi_S(x,t)$  is a solution to the  Cauchy problem for  Dirac equation with the source
\begin{equation}\label{D-2}
i\dot\psi_S(x,t)= D_m \psi_S(x,t) -D^{-1}_m F(\psi(0,t))\delta(x), \quad \psi_S(x,0) = 0.
\end{equation}
The following lemma states well known  local decay  for the free Dirac equation.
\begin{lemma}\label{lemma-decay1} (cf. \cite [Proposition 4.3]{KK2010})
Let $\psi_0\in H^1$. Then $\phi\in C_b(\overline{\R^+}, H^1)$, and $\forall R>0$,
\begin{equation}\label{Uloc}
\Vert \phi(t)\Vert_{H^1(-R,R)}\to 0,\qquad t\to\infty.
\end{equation}
\end{lemma}
Now (\ref{eb}) implies that  
\begin{equation}\label{psi-2-bounds}
\psi_S=\psi(t)-\phi(t)\in C_b(\overline{\R^+}, H^1).
\end{equation}
Due to (\ref{Uloc}) it suffices to prove  (\ref{cal-A}) for $\psi_S$ only.
\subsection*{Complex Fourier-Laplace transform}
Let us analyse the complex Fourier-Laplace transform of $\psi_S(x,t)$:
\begin{equation}\label{FL}
\displaystyle\tilde\psi_S(x,\omega)=\int_0^\infty e^{i\omega t}\psi_S(x,t)\,dt,\quad\omega\in\C^{+},
\end{equation}
where $C^{+}:=\{z\in\C:\;\Im z>0\}$. Due to (\ref{psi-2-bounds}),
$\tilde\psi_S(\cdot,\omega)$ is an $H\sp 1$-valued analytic function of $\omega\in\C\sp{+}$.
\\
Denote $f(t)=F(\psi(0,t)$. Then
equation (\ref{D-2}) for $\psi_S$ with zero initial data implies that
\[
(D_m-\omega)\tilde\psi_S(x,\omega)-D_m^{-1}\delta(x)\tilde f(\omega)=0,\quad \omega\in\C\sp{+},
\]
It is easy to see that 
$$
\tilde\psi_S(x,\omega):=(D_m+\omega)D_m^{-1}\tilde f(\omega)G(x,\omega)
=-\tilde f(\omega)G(x,\omega)-\omega D_m^{-1}\tilde f(\omega)G(x,\omega),
$$
where $G(\cdot,\omega)\in H^1$ is the unique elementary solution to 
\[
G''(x,\omega)+(\omega^2-m^2)G(x,\omega)=\delta(x),\qquad \omega\in\C\sp{+}.
\]
This solution is given by
$\displaystyle G(x,\omega)=\frac{e^{ik(\omega)|x|}}{2ik(\omega)}$, where $k(\omega)$ stands for the analytic function
\begin{equation}\label{def-k}
k(\omega):=\sqrt{\omega\sp 2-m^2},\qquad\Im k(\omega)>0,\qquad\omega\in\C\sp{+},
\end{equation}
which we extend to $\omega\in\Bar{\C\sp{+}}$ by continuity.
Thus,
\begin{equation}\label{tilde-psi-tilde-f}
\tilde\psi_S(x,\omega)=-\tilde f(\omega)\frac{e^{i k(\omega)|x|}}{2ik(\omega)}
-\omega D_m^{-1}\tilde f(\omega)\frac{e^{i k(\omega)|x|}}{2ik(\omega)},\qquad\omega\in\C^{+}.
\end{equation}
Note, that 
$$
D_m^{-2}\frac{e^{i k(\omega)|x|}}{2ik(\omega)}=\frac{1}{2\pi}\int \frac{e^{-ikx}dk}{(k^2+m^2)(-k^2-m^2+\omega^2)}
=\frac{1}{\omega^2}\Big(\frac{e^{-m|x|}}{2m}+\frac{e^{ik(\omega)|x|}}{2ik(\omega)}\Big).
$$
Therefore,
$$
\omega D_m^{-1}\tilde f(\omega)\frac{e^{i k(\omega)|x|}}{2ik(\omega)}
=m\beta \frac{\tilde f(\omega)}{\omega}\Big(\frac{e^{-m|x|}}{2m}+\frac{e^{ik(\omega)|x|}}{2ik(\omega)}\Big)
+\alpha \frac{\tilde f(\omega)}{2\omega}\sgn x(e^{ik(\omega)|x|}-e^{-m|x|}),
$$
and (\ref{tilde-psi-tilde-f}) becomes
\begin{eqnarray}\nonumber
\tilde\psi_S(x,\omega)&=&-\tilde f(\omega)\frac{e^{i k(\omega)|x|}}{2ik(\omega)}
-m\beta \frac{\tilde f(\omega)}{\omega}\Big(\frac{e^{-m|x|}}{2m}+\frac{e^{ik(\omega)|x|}}{2ik(\omega)}\Big)
-\alpha \frac{\tilde f(\omega)}{2\omega}\sgn x(e^{ik(\omega)|x|}-e^{-m|x|})\\
\label{tilde-psi1}
&=&-\Big(I+\beta \,\frac{m+ik(\omega)}{\omega}\Big)\frac{\tilde f(\omega)}{2ik(\omega)}e^{i k(\omega)|x|}
+(\beta-\alpha \sgn x)\tilde f(\omega)\frac{e^{ik(\omega)|x|}-e^{-m|x|}}{2\omega},\qquad\omega\in\C^{+}.
\end{eqnarray}
Here the last term vanishes for $x=0$. Denote $y(t):=\psi_S(0,t)\in C_b(\R)$. Then (\ref{tilde-psi1}) implies
\begin{equation}\label{y-def}
\tilde y(\omega)=\tilde\psi_S(0,\omega)=-\Big(I+\beta \,\frac{m+ik(\omega)}{\omega}\Big)\frac{\tilde f(\omega)}{2ik(\omega)},\quad \omega\in\C^{+}.
\end{equation}
Now  (\ref{tilde-psi1}) becomes
\begin{equation}\label{tilde-psi11}
\tilde\psi_S(x,\omega)=\tilde y(\omega)e^{i k(\omega)|x|}
+(\beta-\alpha \sgn x)\tilde f(\omega)\frac{e^{ik(\omega)|x|}-e^{-m|x|}}{2\omega},\qquad\omega\in\C^{+}.
\end{equation}
Let  us extend $\psi_S(x,t)$ and $f(t)$ by zero for $t<0$. Then
\begin{equation}\label{psi2}
\psi_S\in C_{b}(\R, H^1)
\end{equation}
by (\ref{psi-2-bounds}).
The Fourier transform $\hat\psi_S(\cdot,\omega):=\mathcal{F}_{t\to\omega}[\psi_S(\cdot,t)]$ is a tempered $H\sp 1$-valued distribution of 
$\omega\in\R$. The distribution $\hat\psi_S(\cdot,\omega)$ is the boundary value of the analytic $H^1$-valued function 
$\tilde \psi_S(\cdot,\omega)$, in the following sense:
\begin{equation}\label{bvp1}
\hat\psi_S(\cdot,\omega)=\lim\limits\sb{\varepsilon\to 0+}\tilde\psi_S(\cdot,\omega+i\varepsilon),\qquad\omega\in\R,
\end{equation}
where the convergence holds in the space of tempered distributions $\mathscr{S}'(\R,H^1)$. Indeed,
$$
\tilde\psi_S(\cdot,\omega+i\varepsilon)=\mathcal{F}_{t\to\omega}[\psi_S(\cdot,t)e^{-\varepsilon t}],
$$
and $\psi_S(\cdot,t)e^{-\varepsilon t}\mathop{\longrightarrow}\limits_{\varepsilon\to 0+}\psi_S(\cdot,t)$,
where the convergence holds in $\mathscr{S}'(\R,H^1)$ by (\ref{psi2}).
Therefore, (\ref{bvp1}) holds by the continuity of the Fourier transform $\mathcal{F}_{t\to\omega}$ in $\mathscr{S}'(\R)$.

Similarly to (\ref{bvp1}), the distributions   $\hat f(\omega)$ and $\hat y(\omega)$ of  $\omega\in\R$
are  boundary values of  analytic in $\C^{+}$ functions $\tilde f(\omega)$ and $\tilde y(\omega)$,
$\omega\in\C^{+}$, respectively:
\begin{eqnarray}\nonumber
\hat f(\omega)&=&\lim\limits_{\varepsilon\to 0+}\tilde f(\omega+i\varepsilon),\\
\label{bv}
\hat y(\omega)&=&\lim\limits_{\varepsilon\to 0+}\tilde y(\omega+i\varepsilon)
=-\Big(I+\beta \,\frac{m+ik(\omega)}{\omega}\Big)\frac{\hat f(\omega)}{2ik(\omega)},\quad \omega\in\R,
\end{eqnarray}
 since the function  $f(t)=F(\psi(0,t))$ is bounded for $t\ge 0$ and vanishes for $t<0$.
The convergences hold in the space of tempered distributions $\mathscr{S}'(\R)$.
Let us justify that the representation (\ref{tilde-psi11}) for $\hat\psi_S(x,\omega)$ is also valid when $\omega\in\R$. 
\begin{lemma}\label{prop-uniform}
For any fixed $x\in\R$, 
\begin{equation}\label{p1r}
\hat\psi_S (x,\omega)=\hat y(\omega)e^{i k(\omega)|x|}+(\beta-\alpha \sgn x)\hat f(\omega)\frac{e^{ik(\omega)|x|}-e^{-m|x|}}{2\omega},
\quad\omega\in\R,
\end{equation}
Here the multiplications are understood in the sense of quasimeasures (see \cite[Appendix B]{KK07}). 
\end{lemma}
The proof follows from  (\ref{tilde-psi11})  similarly to  \cite[Proposition 3.1] {KK07}. Namely, the convergence (\ref{bv}) holds in the space of quasimeasures,
while $e^{ik(\omega)|x|}$ and $\frac{e^{ik(\omega)|x|}-e^{-m|x|}}{2\omega}$ are  multiplicators in the space of quasimeasures.
\section{Absolutely continuous spectrum}
\label{abs_con}
Denote
\begin{equation}\label{def-omega-delta}
\Omega_\delta:=(-\infty,-m-\delta)\cup(m+\delta,\infty),\qquad\delta\ge 0.
\end{equation}
Consider the functions
\begin{equation}\label{z-def}
\tilde z^{\,\pm}(\omega):=-\big(I+\frac{m}{\omega}\beta\pm\frac{ik(\omega)}{\omega}\alpha\big) \frac{\tilde f(\omega)}{2ik(\omega)},\quad\omega\in\C^+.
\end{equation}
Let $\hat z^{\,\pm}(\omega)$ be boundary values of $\tilde z^{\,\pm}(\omega)$ on $\Omega_0$:
$$
\hat z^{\,\pm}(\omega)=\lim_{\varepsilon\to 0+}\tilde z^{\,\pm}(\omega+i\varepsilon),\quad\omega\in \Omega_0
$$
Now we rewrite  (\ref{tilde-psi11}) as
\begin{equation}\label{p1r'}
\hat\psi_S (x,\omega)=e^{i k(\omega)|x|}\hat z^{\,\pm}(\omega)
+e^{-m|x|}\big(\pm\alpha-\beta)\frac{\hat f(\omega)}{2\omega},\quad \pm x>0,\quad\omega\in \Omega_0.
\end{equation}
We study the regularity of  $\hat z^{\,\pm}(\omega)$.
Note that the function $\omega k(\omega)$ is positive for $\omega\in\Omega_{0}$.
\begin{proposition}\label{prop-continuity}
The distributions $\hat z^{\,\pm}(\omega)$ are absolutely continuous for $\omega\in\Omega_ 0$, i.e.
and $\hat z^{\,\pm}\in L^1_{loc}(\Omega_0)$. Moreover,
\begin{equation}\label{f2}
\int\sb{\Omega_0}\big(|\hat z^{\,+}(\omega)|^2+|\hat z^{\,-}(\omega)|^2\big) \omega k(\omega)\,d\omega<\infty.
\end{equation}
\end{proposition}
\begin{proof}
We use the arguments of Paley-Wiener type. 
Namely, the Parseval identity and (\ref{psi-2-bounds}) imply that
\begin{equation}\label{PW}
\int\limits_\R
\Vert \tilde\psi_S(\cdot,\omega+i\varepsilon)\Vert_{H^1}^2\,d\omega=2\pi\int\limits_0^\infty e^{-2\varepsilon t}
\Vert \psi_S(\cdot,t)\Vert_{H^1}^2\,dt \le\frac{C}{\varepsilon},\quad \varepsilon>0.
\end{equation}
Then (\ref{PW}) gives
\begin{equation}\label{PW1}
\int\limits_{\R}\Vert \tilde\psi_S(\cdot,\omega+i\varepsilon)\Vert_{H^1}^2\,d\omega \le\frac{C}{\varepsilon},\quad \varepsilon>0.
\end{equation}
Evidently,
$$
\lim_{\varepsilon\to 0+}\varepsilon \Vert e^{-m|x|}\Vert_{H^1}\to 0.
$$
Hence, (\ref{p1r'}) and (\ref{PW1}) results in
\begin{equation}\label{PW-1}
\varepsilon\int_{\Omega_0}\Big(|\tilde z^{\,+}(\omega+i\varepsilon)|^2 \Vert e^{ik(\omega+i\varepsilon)|x|}\Vert_{H^1(0,\infty)}^2
+|\tilde z^{\,-}(\omega+i\varepsilon)|^2 \Vert e^{ik(\omega+i\varepsilon)|x|}\Vert_{H^1(-\infty,0)}^2\Big)\,d\omega\le C,\qquad \varepsilon>0.
\end{equation} 
Here is a crucial observation about the norm of $e^{ik(\omega+i\varepsilon)|x|}$.
\begin{lemma} (cf. \cite [Lemma 3.2] {KK07})
\begin{enumerate}
\item
For $\omega\in\R$,
\begin{equation}\label{wei}
\lim_{\varepsilon\to 0+}\varepsilon \Vert e^{ik(\omega+i\varepsilon)|x|}\Vert_{H^1}^2=n(\omega):=
\left\{\begin{array}{ll}\omega k(\omega), &|\omega|>m \\ 0, &|\omega|<m\end{array} \right.,
\end{equation}
where the norm in $H^1$ is chosen to be
$\Vert\psi\Vert_{H^1}=\left(\Vert \psi'|\Vert_{L^2}^2+m^2\Vert\psi\Vert_{L^2}^2\right)^{1/2}.$
\item
For any $\delta>0$ there exists $\varepsilon_\delta>0$ such that 
\begin{equation}\label{n-half}
\varepsilon \Vert e^{ik(\omega+i\varepsilon)|x|}\Vert_{H^1}^2\ge n(\omega)/2,\quad \omega\in{\Omega_\delta},\quad \varepsilon\in(0,\varepsilon_\delta).
\end{equation}
\end{enumerate}
\end{lemma}
Substituting (\ref{n-half}) into (\ref{PW-1}), we get:
\begin{equation}\label{fin}
\int_{\Omega_\delta}(|\tilde z^{\,+}(\omega+i\varepsilon)|^2+|\tilde z^{\,-}(\omega+i\varepsilon)|^2)\omega k(\omega)\,d\omega\le 2C,
\qquad 0<\varepsilon<\varepsilon\sb\delta,
\end{equation}
with the same constant $C$ as in (\ref{PW-1}), and the region $\Omega\sb\delta$ defined in (\ref{def-omega-delta}).
We conclude that for each $\delta>0$ the set of functions
\[
g^{\pm}_{\delta,\varepsilon}(\omega)=\tilde z^{\,\pm}(\omega+i\varepsilon)|\omega k(\omega)|^{1/2},\qquad\varepsilon\in(0,\varepsilon_\delta),
\]
defined for $\omega\in\Omega_\delta$, is bounded in the Hilbert space $L^2(\Omega_\delta)$, 
and, by the Banach Theorem, is weakly compact. Hence, the convergence of the distributions (\ref{bv})
implies the following weak convergence in the Hilbert space $L^2(\Omega_\delta)$:
\begin{equation}\label{wc}
g^{\pm}_{\delta,\varepsilon}\rightharpoondown g^{\pm}_\delta,\qquad \varepsilon\to 0+,
\end{equation}
where the limit function $g^{\pm}_\delta(\omega)$ coincides with the distribution $\hat z^{\,\pm}(\omega)|\omega k(\omega)|^{1/2}$
restricted onto ${\Omega_\delta}$. It remains to note that the norms of all functions $g^{\pm}_\delta$, $\delta>0$,
are bounded in $L^2(\Omega_\delta)$ by (\ref{fin}), hence (\ref{f2}) follows.
Finally, $\hat z^{\,\pm}(\omega)\in L^1_{loc}(\Omega_0)$ by (\ref{f2}) and the Cauchy-Schwarz inequality.
\end{proof}
\begin{lemma}\label{Lem-ek}
\begin{equation}\label{f3}
\int_{\Omega_0}|\hat f(\omega)|^2 \frac{k(\omega)}{\omega}\,d\omega<\infty.
\end{equation}
\end{lemma}
\begin{proof}
Denote the $2\times 2$ matrix $A^{\pm}(\omega)=I+\frac{m}{\omega}\beta\pm\frac{ik(\omega)}{\omega}\alpha$. 
Then $\hat z^{\,\pm}(\omega)=-A^{\pm}(\omega)\frac{\hat f(\omega)}{2ik(\omega)}$. 
For any $\omega\in\Omega_0$, the matrix  $A^{\pm}(\omega)$ has two eigenvalues: $\lambda=0$ and $\lambda=2$ since
$$
{\rm det} (A^{\pm}-\lambda I)=\left |\begin{array}{cc} (1+\frac{m}{\omega})-\lambda & \pm\frac{ik(\omega)}{\omega}\\\\
\mp\frac{ik(\omega)}{\omega} & (1-\frac{m}{\omega})-\lambda\end{array}\right|
=-2\lambda+\lambda^2=\lambda(\lambda -2).
$$
The unit eigenvectors $\nu^{\pm}(\omega)$ of operators $A^{\pm}(\omega)$ corresponding to the eigenvalue $\lambda=2$ read
\begin{equation}\label{nupm}
\nu^{\pm}(\omega)=\big(\sqrt{\frac{\omega+m}{2\omega}},~\mp i\sqrt{\frac{\omega-m}{2\omega}}\big).
\end{equation}
Denote $g(\omega):=\frac{\hat f(\omega)}{2ik(\omega)}$, 
and   $a^{\pm}(\omega):=\langle g(\omega),\nu^{\pm}(\omega)\rangle$. 
Then $\hat z^{\,\pm}(\omega)=-A^{\pm}(\omega)g(\omega)=-2a^{\pm}(\omega)\nu^{\pm}(\omega)$, and hence
$$
a^{\pm}(\omega)=\langle g(\omega),\nu^{\pm}(\omega)\rangle=-\frac 12\langle \hat z^{\,\pm}(\omega), \nu^{\pm}(\omega)\rangle.
$$
Taking into account (\ref{nupm}), we get the system of equations for $g(\omega)=(g_1(\omega),g_2(\omega))$
\begin{equation}\label{sys0}
\left \{\begin{array}{cc}
g_1(\omega)\sqrt{\frac{\omega+m}{2\omega}}+ig_2(\omega)\sqrt{\frac{\omega-m}{2\omega}}
=-\frac 12\langle \hat z^{\,+}(\omega), \nu^{+}(\omega)\rangle
\\\\
g_1(\omega)\sqrt{\frac{\omega+m}{2\omega}}-ig_2(\omega)\sqrt{\frac{\omega-m}{2\omega}}
=-\frac 12\langle \hat z^{\,-}(\omega), \nu^{-}(\omega)\rangle
\end{array}\right|
\end{equation}
Therefore
$$
\left (\begin{array}{cc} \hat f_1(\omega)\\ \hat f_2(\omega)\end{array}\right)
=2ik(\omega)\left (\begin{array}{cc} g_1(\omega)\\ g_2(\omega)\end{array}\right)=-|\omega|
\left (\begin{array}{cc} i\sqrt{\frac{\omega-m}{2\omega}} &  i\sqrt{\frac{\omega-m}{2\omega}} \\\\
\sqrt{\frac{\omega+m}{2\omega}}& -\sqrt{\frac{\omega+m}{2\omega}}\end{array}\right)
\left (\begin{array}{cc} \langle \hat z^{\,+}(\omega), \nu^{+}(\omega)\rangle\\ \langle \hat z^{\,-}(\omega), \nu^{-}(\omega)\rangle\end{array}\right)
$$
Hence,
$$
|\hat f(\omega)|^2\le C|\omega|^2(|\hat z^{\,+}(\omega)|^2+|\hat z^{\,-}(\omega)|^2).
$$
Now (\ref{f3}) follows from (\ref{f2}).
\end{proof}
\section{Further decomposition of  solutions}
\label{decom-2}
Denote
\begin{equation}\label{dpo}
\hat f_{d}(\omega):=\left\{\begin{array}{cc} \hat f(\omega), &\quad\omega\in\Omega_0\\
0,  & \quad\omega\in\R\setminus\Omega_0
\end{array}\right.
\end{equation}
and set
\begin{equation}\label{ftd}
\hat\psi_{d}(x,\omega)=e^{i k(\omega)|x|}\hat z_d(x,\omega)+e^{-m|x|}\big(\alpha\, \sgn x-\beta\big)\frac{\hat f_d(\omega)}{2\omega},\quad\omega\in\R,
\end{equation}
where
\begin{equation}\label{zx}
\hat z_d(x,\omega)=-\big(I+\frac{m}{\omega}\beta+\frac{ik(\omega)}{\omega}\alpha\sgn x\big) \frac{\hat f_d(\omega)}{2ik(\omega)}
=-\big(I+\frac{m}{\omega}\beta\pm\frac{ik(\omega)}{\omega}\alpha\big) \frac{\hat f_d(\omega)}{2ik(\omega)}
=\hat z_d^{\,\pm}(\omega),\quad \pm x>0
\end{equation}
Consider
\begin{equation}\label{dd}
\psi_{d}(x,t):=\frac 1{2\pi}\int_{\R}\hat\psi_{d}(x,\omega)e^{-i\omega t}\,d\omega,\qquad x\in\R,\quad t\in\R.
\end{equation} 
We will show that $\psi_d(x,t)$ is a dispersive component of the solution $\psi(x,t)$, in the following sense.
\begin{proposition}\label{prop-decay-psi-d}
(\it i) $\psi_{d}(\cdot,t)$ is a bounded continuous $H^1$-valued function:
\begin{equation}\label{bound-psi-d}
  \psi_{d}(\cdot,t)\in C_{b}(\R,H^1).
\end{equation}
(\it ii) The local energy decay holds for $\psi_{d}(\cdot,t)$: for any $R>0$,
\begin{equation}\label{decay-psi-d}
\Vert\psi_{d}(\cdot,t)\Vert_{H^1(-R,R)}\to 0, \qquad t\to\infty.
\end{equation}
\end{proposition}
\begin{proof}
We split $\psi_{d}(x,t)$ as $\psi_{d}(x,t)=\varphi_{d}(x,t)+\chi_{d}(x,t)$, where
\begin{eqnarray}\label{splitpsiS}
\varphi_{d}(x,t)=\frac 1{2\pi}\int_{\R}e^{-i\omega t}e^{i k(\omega)|x|}\hat z_d(x,\omega)\,d\omega,\qquad
\chi_{d}(x,t)=\frac 1{2\pi}e^{-m|x|}\big(\alpha\sgn x-\beta\big)\int_{\R}e^{-i\omega t}\frac{\hat f_d(\omega)}{2\omega}\,d\omega.
\end{eqnarray}
First,  consider $\chi_{d}(x,t)$. Note that
\begin{equation}\label{ekn}
\int\limits_{\R}\big|\frac{\hat f_d(\omega)}{\omega}\big|\,d\omega=
\int\limits_{\Omega_{0}}\big|\frac{\hat f_d(\omega)}{\sqrt {\omega k(\omega)}}\sqrt{\frac{k(\omega)}{\omega}}\big|\,d\omega
\le \Big(\int\limits_{\Omega_{0}}|\hat f(\omega)|^2\frac{k(\omega)}{\omega}\,d\omega\Big)^{1/2}
 \Big(\int\limits_{\Omega_{0}}\frac{d\omega}{\omega \sqrt{\omega^2-m^2})}\Big)^{1/2}<\infty
\end{equation}
by Lemma \ref{Lem-ek}. Hence, 
\begin{equation}\label{psid2}
\chi_{d}\in C_{b}(\R, H^1(\R\setminus 0)),
\end{equation}
Moreover,
\begin{equation}\label{dpsid2}
\Vert \chi_{d}(\cdot,t)\Vert_{H^1(\R\setminus 0)}\to 0,\qquad t\to\infty
\end{equation}
by Riemann-Lebesgue Theorem. Now consider $\varphi_{d}(x,t)$. 
Changing the variable  $\omega\to k(\omega)=\sqrt{\omega^2-m^2}$, we rewrite  $\varphi_{d}(x,t)$ as follows:
\begin{equation}\label{ppmr}
\varphi_{d}(x,t)=\frac 1{2\pi}\int_{\R}\hat z_d(x,\omega(k))e^{-i\omega(k) t}e^{i k|x|}\frac{k\,dk}{\omega(k)}.
\end{equation} 
Here $\omega(k)=\sqrt{k^2+m^2}$ is the branch analytic for $\Im k>0$ and continuous for  $\Im k\ge 0$.
Note that the function $\omega(k)$, $k\in\R\backslash 0$, is the inverse function to
$k(\omega)$ defined on $\Bar{\C\sp{+}}$ (see (\ref{def-k})) and restricted onto $\Omega_0$.
 Let us introduce the functions
\begin{eqnarray}\nonumber
\varphi^\pm(x,t)=\frac 1{2\pi}\int_\R \hat z_d^{\,\pm}(\omega(k))e^{\pm ikx}e^{-i\omega(k) t}\frac{kdk}{\omega(k)}, \quad x\in\R,\quad t\ge 0.
\end{eqnarray}
Both functions $\varphi^{\pm}(x,t)$ are solutions to the free Dirac equation (\ref{D-1}) on the whole real line (see Appendix B).  Moreover,
\begin{equation}\label{ppm2}
\partial_x\varphi^{\pm}(x,t)\!:=\frac 1{2\pi}\int_\R \pm ik\, \hat z_d^{\,\pm}(\omega(k)) e^{\pm ikx}e^{-i\omega(k) t}\frac{kdk}{\omega(k)}, \quad x\in\R,\quad t\ge 0.
\end{equation}
Hence, the Parseval identity implies
\begin{eqnarray}
\Vert \varphi^{\pm}(\cdot,0)\Vert_{H^1}^2&=&\int_{\R} (m^2+k^2)|\hat z_d^{\,\pm}(\omega(k))|^2\frac{k^2}{\omega^2(k)}\,dk\\
\nonumber
&=&\int_{\Omega_0} \omega^2|\hat z^{\,\pm}(\omega)|^2\frac{k(\omega)}{\omega}\,d\omega
=\int_{\Omega\sb 0}|\hat z^{\,\pm}(\omega)|^2\omega k(\omega)\,d\omega<\infty.
\nonumber
\end{eqnarray}
by (\ref{fin}). Hence, both $\varphi^{-}$ and $\varphi^{+}$ are bounded continuous
$H^1$-valued functions:
\begin{equation}\label{psipm}
\varphi^{\pm}\in C_{b}(\R, H^1),
\end{equation}
and for any $R>0$
\begin{equation}\label{psi-pm-local-decay}
\Vert \varphi^{\pm}(\cdot,t)\Vert_{H^1(-R,R)}\to 0,\qquad t\to\infty
\end{equation}
by  Lemma~\ref{lemma-decay1}. The function $\varphi_{d}(x,t)$ coincides with $\varphi^{+}(x,t)$ for $x\ge 0$ and with $\varphi^{-}(x,t)$ for $x\le 0$:
\begin{equation}\label{psid1}
\varphi_{d}(x,t)=\varphi^{\pm}(x,t),\qquad \pm x\ge 0.
\end{equation}
It remains to note that  $\psi_{d}(x,t)=\varphi_{d}(x,t)+\chi_{d}(x,t)$ has no jump at $x=0$ and therefore $\partial_x\psi_{d}(x,t)$ is square-integrable over the whole $x$-axis.
Hence,
$$
\Vert\psi_{d}(t)\Vert_{H^1}^2=\Vert\psi_{d}(t)\Vert_{H^1(\R^-)}^2+\Vert\psi_{d}(t)\Vert_{H^1(\R^+)}^2.
$$
Finally,  (\ref{bound-psi-d})  follows from (\ref{psipm}) and (\ref{psid2}), and  (\ref{decay-psi-d}) 
follows from (\ref{psi-pm-local-decay}) and (\ref{dpsid2}).
\end{proof}
Denote $y_d(t)=\psi_d(0,t)\in C_b(|\R)$. Formulas (\ref{ftd}) and (\ref{zx}) imply
\begin{equation}\label{yd}
\hat y_d(\omega)=\hat\psi_d(0,\omega)=-\big(I+\frac{m+ik(\omega)}{\omega}\beta\big) \frac{\hat f_{d}(\omega)}{2ik(\omega)},, \qquad\omega\in\R,
\end{equation}
and (\ref{ftd}) becomes
\begin{equation}\label{ftd0}
\hat\psi_{d}(x,\omega)=\hat y_de^{-\varkappa(\omega)|x|}
+(\beta-\alpha \sgn x)\hat f(\omega)\frac{e^{ik(\omega)|x|}-e^{-m|x|}}{2\omega}, \qquad\omega\in\R.
\end{equation}
\section{Bound component}
\label{sect-bound}
\subsection*{Spectral representation}
We introduce the bound component of the solution $\psi(x,t)$ by
\begin{equation}\label{bb}
\psi_{b}(x,t)=\psi_S(x,t)-\psi_{d}(x,t),\ \ x\in\R,\ \ t\in\R.
\end{equation}
Then (\ref{psi2}) and (\ref{bound-psi-d}) imply that
\begin{equation}\label{ebb}
\psi_{b}\in C_{b}(\R,H^1).
\end{equation}
In particular, $y_b(t):=\psi_{b}(0,t)=\psi_S(0,t)-\psi_{d}(0,t)\in C_b(\R)$. Hence,  $\hat y_b(\omega):=\hat \psi_{b}(0,\omega)$ is a quasimeasure.
Moreover, formulas (\ref{bv}) and (\ref{yd}) yield  
\begin{equation}\label{yb}
\hat y_b(\omega)=\hat y(\omega)-\hat y_d(\omega)
=-\big(I+\frac{m+ik(\omega)}{\omega}\beta\big) \frac{\hat f_{b}(\omega)}{2ik(\omega)}.
\end{equation}
 Here we denote
\begin{equation}\label{fb}
\hat f_{b}(\omega):=\hat f(\omega)-\hat f_{d}(\omega).
\end{equation}
Further, (\ref{dpo}) implies, that
\begin{equation}\label{ftbs}
\supp \hat y_b(\omega)=\supp \hat\psi_b(0,\omega)\subset [-m,m].
\end{equation}
Denote
\begin{equation}\label{def-kappa}
\varkappa(\omega):=-i k(\omega)=\sqrt{m^2-\omega^2},\qquad\Re\varkappa(\omega)\ge 0\quad{\rm for}\quad\Im\omega\ge 0,
\end{equation}
where $k(\omega)$ was introduced in (\ref{def-k}). Let us note that $\varkappa(\omega)>0$ for $\omega\in (-m,m)$. Now  we rewrite (\ref{yb}) as
\begin{equation}\label{psib0}
\hat y_b(\omega)=\big(I+\frac{m-\varkappa(\omega)}{\omega}\beta\big) \frac{\hat f_{b}(\omega)}{2\varkappa(\omega)}=\sigma(\omega)\hat f_b(\omega),
\end{equation}
where
$$
\sigma(\omega)=\frac{1}{2\varkappa(\omega)}\left(\begin{array}{cc}
1+\frac{m-\varkappa(\omega)}{\omega} & 0\\
0 & 1-\frac{m-\varkappa(\omega)}{\omega}\end{array}\right)
$$
Hence
\begin{equation}\label{hatf}
\hat f_b(\omega)=\sigma^{-1}(\omega)\hat y_b(\omega),\qquad\sigma^{-1}(\omega)=\left(\begin{array}{cc}
\varkappa(\omega)+m-\omega & 0\\
0 & \varkappa(\omega)+m+\omega \end{array}\right).
\end{equation}
Now (\ref{p1r}), (\ref{ftd0}), (\ref{bb}) and (\ref{hatf}) imply the multiplicative relation
\begin{equation}\label{ftb0}
\hat\psi_{b}(x,\omega)=\hat y_be^{-\varkappa(\omega)|x|}
+\hat h_b(x,\omega)\frac{e^{-\varkappa(\omega)|x|}-e^{-m|x|}}{2\omega}, \qquad\omega\in\R,
\end{equation}
where we denote 
\begin{equation}\label{hath}
\hat h_b(x,\omega)=(\beta-\alpha \sgn x)\sigma^{-1}(\omega)\hat y_b(\omega).
\end{equation}
From  (\ref{hatf}) and (\ref{hath}) it follows  that $\hat h_b(x,\omega)$ for any fixed $x\in\R\setminus 0$ is a quasimeasure with the support $\supp\hat h_b(x,\omega)\subset[-m,m]$. 
Moreover, $e^{-\varkappa(\omega)|x|}$ and $(e^{-\varkappa(\omega)|x|}-e^{-m|x|})/\omega$ are multiplicators.
Hence, function $\hat\psi_{b}(x,\omega)$ is quasimeasure for any fixed $x\in\R$ with supports in $[-m,m]$.
Finally,
\begin{equation}\label{bdef}
\psi_{b}(x,t)=\frac 1{2\pi}\langle \hat\psi_{b}(x,\omega), e^{-i\omega t}\rangle,\quad x\in\R,\quad t\in\R,
\end{equation} 
where $\langle \cdot,\cdot\rangle$ is an extension of the scalar product
$\langle f,g\rangle=\int f(\omega)\ov g(\omega)d\omega$.
\subsection*{Compactness}
We are going to prove a compactness of the set of translations of the bound component, $\{\psi_{b,n}(x,s+t)\sothat s\ge 0\}$, $n=1,2$.
\begin{lemma} \label{L1}
(i) The function $\psi_{b}(x,t)$  is smooth for $x\ne 0$ and $t\in\R$. Moreover, for any fixed $x\not=0$, $t\in\R$, and 
 any nonnegative  integers $j,k$,   the following representation holds
\begin{equation}\label{ek}
\partial_x^j \partial_t^k \psi_{b}(x,t)=\frac 1{2\pi}\langle \Lambda_j(x,\omega), (-i\omega)^k e^{-i\omega t}\rangle,
\end{equation} 
where
\begin{eqnarray}\nonumber
\Lambda_j(x,\omega)&=&\big(\!\!-\!\varkappa(\omega)\sgn x\big)^j e^{-\varkappa(\omega)|x|}\hat y_b(\omega)\\
\nonumber
&+&\hat h_b(x,\omega)\big[\big(\!\!-\!\varkappa(\omega)\sgn x\big)^j\,\frac{e^{-\varkappa(\omega)|x|}-e^{-m|x|}}{2\omega}
+\frac{\varkappa^j(\omega)-m^j}{\omega}(-\sgn x)^je^{-m|x|}\big].
\end{eqnarray} 
(ii)      There is a constant $C_{j,k}>0$ so that
\begin{equation}\label{bqda}
\sup\limits_{x\not =0}\,\,\sup\limits_{t\in\R}\big|\partial_x^j \partial_t^k\psi_{b}(x,t)\big|\le C_{j,k}.
\end{equation}
\end{lemma}
The lemma follows similarly Proposition 4.1 from \cite {KK07}, since the factors
$e^{-\varkappa(\omega)|x|}\zeta(\omega)$, $\frac{e^{-\varkappa(\omega)|x|}-e^{-m|x|}}{2\omega}\zeta(\omega)$,  and
$\frac{\varkappa^j(\omega)-m^j}{\omega}\zeta(\omega)$
are multiplicators in the space of quasimeasures.  Here $\zeta(\omega)\in C_{0}^\infty(\R)$ is any cutoff function satisfying
$$
\zeta|_{[-m-1,m+1]}=1.
$$
\begin{corollary}\label{coco}
By the Ascoli-Arzel\`a Theorem, for any sequence $s_{l}\to\infty$ there exists a subsequence (which we also denote by $s_l$)
such that for any nonnegative integers $j$ and $k$,
\begin{equation}\label{olpd}
\partial_x^j \partial_t^k \psi\sb{b}(x,s_{l}+t)\to \partial_x^j \partial_t^k \gamma (x,t),\qquad x\ne 0,\quad t\in\R.
\end{equation}
for some $\gamma\in C_{b}(\R,H\sp 1)$. The convergence in {\rm (\ref{olpd})} is uniform in $x$ and $t$ as long as $|x|+|t|\le R$, for any $R>0$.
\end{corollary}
We call {\it omega-limit trajectory} any function $\gamma(x,t)$ that can appear as a limit in (\ref{olpd}).
Previous analysis demonstrates that the long-time asymptotics of the solution $\psi(x,t)$ in $H^1_{loc}$
depends only on the bound component $\psi_{b}(x,t)$.
By Corollary \ref{coco}, to conclude the proof of Theorem~\ref{main-theorem}, it suffices to check that every omega-limit trajectory
belongs to the set of solitary waves; that is,
\begin{equation}\label{eidd}
\gamma(x,t)=\psi_{\Omega_{+}}(x,t)+iD_m^{-1}\dot\psi_{\Omega_{+}}(x,t) \qquad x,\,t\in\R,
\end{equation}
where $\psi_{\Omega^{+}}(x,t)=(e^{-i\omega_1^{+}t}\phi_{\omega_1^+}(x), e^{-i\omega_2^{+}t}\phi_{\omega_2^+}(x))$
with some $\omega_1^{+},\omega_2^+\in[-m,m]$.
\subsection*{Spectral identity for omega-limit trajectories}
Here we study the time spectrum of the omega-limit trajectories.
\begin{definition}
Let $\mu$ be a tempered distribution. By $\Spec \mu$ we denote the support of its Fourier transform:
\[
\Spec \mu:=\supp\tilde \mu.
\]
\end{definition}
\begin{proposition}\label{prop-beta}
\begin{enumerate}
\item
For any omega-limit trajectory $\gamma(x,t)$, the following spectral representation holds:
\begin{equation}\label{ber}
\gamma(x,t)=\frac 1{2\pi}\langle \hat p(\omega)e\sp{-\varkappa(\omega)|x|},e\sp{-i\omega t}\rangle
+\frac 1{2\pi}\langle\hat q(x,\omega)\frac{e^{-\varkappa(\omega)|x|}-e^{-m|x|}}{2\omega},e\sp{-i\omega t}\rangle,
\qquad x\in\R,\qquad t\in\R,
\end{equation}
where $\hat p(\omega)$  and $\hat q(x,\omega)=(\beta-\alpha \sgn x)\sigma^{-1}(\omega)\hat p(\omega)$ are quasimeasures for all $x\in\R$, and
\begin{equation}\label{spec-gamma}
\supp \hat p\subset [-m,m],\qquad \supp \hat q(x)\subset [-m,m].
\end{equation}
\item
The following bound holds:
\begin{equation}\label{ebbe}
\sup\limits\sb{t\in\R}
\Vert \gamma(\cdot,t)\Vert_{H^1}<\infty.
\end{equation}
\end{enumerate}
\end{proposition}
Note that, according to (\ref{ber}), $\hat p(\omega)$ is the Fourier transform of the function $p(t):= \gamma(0,t)$, $t\in\R$.
\begin{proof}
Formula (\ref{ftb0}) and representation (\ref{bdef}) imply that
\begin{equation}\label{bqdwi}
\psi_{b}(x,s_{l}+t)=\frac{1}{2\pi}\langle \hat y_b(\omega)e^{ -\varkappa(\omega)|x|}e^{-i\omega s_{l}},e^{-i\omega t}\rangle
+\frac{1}{2\pi}\langle\hat h_b(x,\omega)\frac{e^{-\varkappa(\omega)|x|}-e^{-m|x|}}{2\omega}e^{-i\omega s_{l}},e^{-i\omega t}\rangle,
\quad x\ne 0,\quad t\in\R.
\end{equation}
Further, the convergence  (\ref{olpd}) and the bound (\ref{bqda}) with  $j=k=0$ imply that
\begin{equation}\label{dztA}
y_b(s_{l}+t)\to p(t),\qquad s_{l}\to\infty,
\end{equation}
where $p\in C_b(\R)$.  The convergence is uniform on $[-T,T]$ for any $T>0$. Hence, 
\begin{equation}\label{dztAd}
\hat y_b(\omega)e^{-i\omega s_{l}}\to \hat p(\omega), \qquad s_{l}\to\infty.
\end{equation}
in the space of  quasimeasures. Therefore,
\begin{equation}\label{flA}
\hat y_b(\omega)e^{ -\varkappa(\omega)|x|}e^{-i\omega s_{l}}\to \hat p(\omega)e^{-\varkappa(\omega)|x|}, \qquad s_{l}\to\infty.
\end{equation}
in the space of  quasimeasures. Similarly,
\begin{eqnarray}\nonumber
&&\hat h_b(x,\omega)\frac{e^{-\varkappa(\omega)|x|}-e^{-m|x|}}{2\omega}e^{-i\omega s_{l}}
=(\beta-\alpha\sgn x)\sigma^{-1}(\omega)\hat y_b(\omega)\frac{e^{-\varkappa(\omega)|x|}-e^{-m|x|}}{2\omega}e^{-i\omega s_{l}}\\
\label{flA1}
&&\to (\beta-\alpha \sgn x)\sigma^{-1}(\omega)\hat p(\omega)\frac{e^{-\varkappa(\omega)|x|}-e^{-m|x|}}{2\omega}
=\hat q(x,\omega) \frac{e^{-\varkappa(\omega)|x|}-e^{-m|x|}}{2\omega},\qquad s_{l}\to\infty.
\end{eqnarray}
Hence, the representation (\ref{ber}) follows from (\ref{bqdwi}), (\ref{flA}) and (\ref{flA1}); and (\ref{spec-gamma}) follows from (\ref{ftbs}).
Finally, the bound (\ref{ebbe}) follows from (\ref{ebb}) and (\ref{olpd}).
\end{proof}
The relation (\ref{ber}) implies the basic spectral identity:
\begin{corollary}\label{corol}
For any omega-limit trajectory $\gamma(x,t)$,
\begin{equation}\label{SiI}
\Spec \gamma(x,\cdot)=\Spec p,\qquad x\in\R.
\end{equation}
\end{corollary}
\section{Nonlinear spectral analysis}
\label{sect-spectral}
Here we will derive (\ref{eidd}) from the following identity:
\begin{equation}\label{eide}
p_j(t)=C_je^{-i\omega_j^{+}t},\qquad j=1,2,\qquad t\in\R,
\end{equation}
which will be proved in three steps.
\subsection*{Step 1}
The identity (\ref{eide}) is equivalent to
$\hat p_j(\omega)\sim\delta(\omega-\omega_j^{+})$, so we start with an investigation of $\Spec p_j:=\supp \hat p_j$.
\begin{lemma}
For omega-limit trajectories the following spectral inclusion holds:
\begin{equation}\label{Si}
\Spec F_j(p_j(\cdot))\subset \Spec p_j,\qquad j=1,2.
\end{equation}
\end{lemma}
\begin{proof}
The convergence  (\ref{olpd}), Lemma~\ref{lemma-decay1} and Proposition~\ref{prop-decay-psi-d}~({\it ii})
imply that the limiting trajectory $\gamma(x,t)$ is a solution to equation (\ref{KG-0}):
\begin{equation}\label{D-beta}
i\dot\gamma(x,t)=D_m\gamma(x,t)-D_m^{-1}\delta(x)F(\gamma(0,t)), \qquad (x,t)\in\R^2.
\end{equation}
Applying  to both side operator $D_m$, we get
$$
iD_m\dot\gamma(x,t)=D_m^2\gamma(x,t)-\delta(x)F(\gamma(0,t)), \qquad (x,t)\in\R^2.
$$
Since $\gamma(x,t)$ is smooth function for $x\le 0$ and $x\ge 0$, we get the following algebraic identities :
\begin{equation}\label{AI}
\gamma'_j(0+,t)-\gamma'_j(0-,t)=-F_j(p_j(t)), \quad t\in\R,\quad j=1,2.
\end{equation}
The identities imply the spectral inclusion
\begin{equation}\label{spectral-inclusion}
\Spec F_j(p_j(\cdot))\subset\Spec \gamma'_j(0+,\cdot)\cup\Spec \gamma'_j(0-,\cdot).
\end{equation}
On the other hand,
$\Spec \gamma'_j(0+,\cdot)\cup\Spec \gamma'_k(0-,\cdot)\subset\Spec p_j$
by (\ref{SiI}).
Therefore, (\ref{spectral-inclusion}) implies (\ref{Si}).
\end{proof}
\subsection*{Step 2}
\begin{proposition}\label{pTN}
For any omega-limit trajectory, the following identity holds:
\begin{equation}\label{C}
|p_j(t)|={\rm const},\qquad j=1,2, \qquad t\in\R.
\end{equation}
\end{proposition}
\begin{proof}
We are going to show that (\ref{C}) follows from the key spectral relations (\ref{spec-gamma}), (\ref{Si}).
Our main assumption (\ref{f-is-such}) implies that the function $F_j(t):=F_j(p_j(t))$ admits the representation (cf. (\ref{def-a}))
\begin{equation}\label{frep}
F_j(t)=a_j(t)p_j(t),\quad j=1,2,
\end{equation}
where, according to (\ref{f-is-such}),
\begin{equation}\label{arep}
a_j(t)=-\sum\limits_{n=1}^{N_j}  2n_j u_{n,j}|p_j(t)|^{2n-2},\qquad N_j\ge 2;\quad u_{N_j,j} > 0.
\end{equation}
Both functions $p_j(t)$ and $a_j(t)$ are bounded continuous functions in $\R$ by Proposition~\ref{prop-beta}~({\it ii}).
Hence, $p_j(t)$ and $a_j(t)$ are tempered distributions.
Furthermore, $\hat p_j$ and $\hat{\overline p}_j$ have the supports contained in $[-m,m]$ by (\ref{spec-gamma}).
Hence, $a_j$ also has a bounded support since it is a sum of convolutions of finitely many $\hat p_j$ and $\hat{\overline p}_j$ by (\ref{arep}).
Then the relation (\ref{frep}) translates into a convolution in the Fourier space,
$\hat F_j=\hat a_j\ast\hat p_j/(2\pi),$ and the spectral inclusion (\ref{Si}) takes the following form:
\begin{equation}\label{Si-ast}
\supp\hat F_j= \supp\,\hat a_j\ast\hat p_j\subset\supp\hat p_j.
\end{equation}
Let us  denote ${\bf F}_j=\supp \hat F_j$, ${\bf A}_j=\supp \hat a_j$, and ${\bf P}_j=\supp\hat p_j$. Then the spectral inclusion (\ref{Si-ast}) reads as
\begin{equation}\label{Si-ast-r}
{\bf F}_j\subset {\bf P}_j.
\end{equation}
On the other hand, it is well-known that $\supp\hat a_j\ast\hat p_j\subset\supp\hat a_j+\supp \hat p_j$,
or ${\bf F}_j \subset {\bf A}_j+{\bf P}_j.$
Moreover, the Titchmarsh convolution theorem (see  \cite[Theorem 4.3.3]{Hor90}) imply that
\begin{equation}\label{msc}
\inf {\bf F}_j=\inf {\bf A}_j+\inf {\bf P}_j,\qquad \sup {\bf F}_j=\sup {\bf A}_j+\sup {\bf P}_j.
\end{equation}
Now  (\ref{Si-ast-r}) and (\ref{msc}) result in
\begin{equation}\label{msci}
\inf {\bf F}_j=\inf {\bf A}_j+\inf {\bf P}_j\ge\inf {\bf P}_j,\qquad \sup {\bf F}_j=\sup {\bf A}_j+\sup {\bf P}_j\le \sup {\bf P}_j,
\end{equation}
so that $\inf {\bf A}_j\ge 0\ge \sup {\bf A}_j$. Thus, we conclude that $\supp\hat a_j={\bf A}_j\subset\{0\}$, therefore the distribution 
$\hat a_j(\omega)$ is a finite linear combination of $\delta(\omega)$ and its derivatives.
Then $a_j(t)$ are  polynomial in $t$; since $a_j(t)$ is bounded by Proposition~\ref{prop-beta}~({\it ii}),
we conclude that $a_j(t)$ is constant. Now the relation (\ref{C}) follows since $a_j(t)$ is a polynomial  in
$|p_j(t)|$, and its degree is strictly positive by (\ref{arep}).
\end{proof}
\subsection*{Step 3}
Now the same Titchmarsh arguments imply that $P_j:=\Spec p_j$ is a point $\omega_j^{+}\in[-m,m]$. Indeed,
(\ref{C}) means that $p_j(t) \overline p_j(t)\equiv C_j$, hence in the Fourier transform $\hat p_j \ast\hat{\overline p_j}=2\pi C_j\delta(\omega)$.
Therefore, if $p_j$ is not identically zero, the Titchmarsh Theorem implies that
\[
0=\sup P_j+\sup(-P_j)=\sup P_j-\inf P_j.
\]
Hence $\inf P_j=\sup P_j$ and therefore $P_j=\omega_k^{+}\in [-m,m]$, so that $\hat p_j(\omega)$ is a finite linear combination
of $\delta(\omega-\omega_j^{+})$ and its derivatives. As the matter of fact, the derivatives could not be present
because of the boundedness of $p_j(t)=\gamma_j(0,t)$ that follows from Proposition~\ref{prop-beta}~({\it ii}).
Thus, $\hat p_j\sim\delta(\omega-\omega_j^{+})$, which implies (\ref{eide}).
\subsubsection*{Conclusion of the proof of Theorem~\ref{main-theorem}}
According to (\ref{eide}) and (\ref{flA1})
$$
\hat q(\omega,x)=(\beta-\alpha\sgn x)\sigma^{-1}(\omega)\hat p(\omega)
=2\pi(\beta-\alpha\sgn x)\left(\begin{array}{cc}
C_1(\varkappa_1^++m-\omega_1^+ )\delta(\omega-\omega_1^{+})\\ 
C_2( \varkappa_2^++\omega_2^++m)\delta(\omega-\omega_2^{+})\end{array}\right),
$$
where
$$
\varkappa_j^+=\sqrt{m^2-(\omega_j^+)^2}.
$$
Then the representation (\ref{ber}) implies 
\begin{equation}\label{gamma12}
\left \{\begin{array}{cc}\gamma_1(x,t)=C_1e^{-i\omega_1^+ t}\Big(\!e^{-\varkappa_1^+|x|}+(\varkappa_1^+\!\!+m-\!\omega_1^+)
\frac{e^{-\varkappa_1^+|x|}-e^{-m|x|}}{2\omega_1^+}\Big)
-C_2e^{-i\omega_2^+ t}\sgn x( \varkappa_2^+\!\!+m+\!\omega_2^+)\frac{e^{-\varkappa_2^+|x|}\!-e^{-m|x|}}{2\omega_2^+}\\\\
\gamma_2(x,t)=C_2e^{-i\omega_2^+ t}\Big(\!e^{-\varkappa_2^+|x|}-(\varkappa_2^+\!\!+m+\!\omega_2^+)
\frac{e^{-\varkappa_2^+|x|}- e^{-m|x|}}{2\omega_2^+}\Big)
+C_1e^{-i\omega_1^+ t}\sgn x(\varkappa_1^+\!\!+m-\!\omega_1^+)\frac{e^{-\varkappa_1^+|x|}\!-e^{-m|x|}}{2\omega_1^+}
\end{array}\right|
\end{equation}
After simple evaluation, (\ref{gamma12}) becomes
$$
\left \{\begin{array}{cc}
\gamma_1(x,t)=C_1'\Big(e^{-\varkappa_1^+|x|}
+\frac{m e^{-\varkappa_1^+|x|}-\varkappa_1e^{-m|x|}}{\omega_1^+}\Big)e^{-i\omega_1^+ t}
-C_2'\varkappa_2^+\sgn x\frac{e^{-\varkappa_2^+|x|}-e^{-m|x|}}{\omega_2^+} e^{-i\omega_2^+ t}\\\\
\gamma_2(x,t)=C'_1\varkappa_1^+\sgn x\frac{e^{-\varkappa_1^+|x|}-e^{-m|x|}}{\omega_1^+}e^{-i\omega_1^+ t}
+C_2'\Big(e^{-\varkappa_2^+|x|}-\frac{m e^{-\varkappa_2^+|x|}-\varkappa_2 e^{-m|x|}}{\omega_2^+}\Big)e^{-i\omega_2^+ t}
\end{array}\right|
$$
where we denote
$$
C_1'=C_1\frac{\varkappa_1^++m-\omega_1^+}{2\varkappa_1^+},\quad C_2'=C_2\frac{\varkappa_2^++m+\omega_2^+}{2\varkappa_2^+}.
$$
Therefore, $\gamma(x,t)$ is a solitary wave  (\ref{psiOmsol}).
Due to Lemma \ref{lemma-decay1}  and Proposition \ref{decay-psi-d}
it remains to prove that \begin{equation}\label{cal-B}
\lim_{t\to\infty}
{\rm dist}_{H^1_{loc}}(\psi_b(t),{\bf S})=0. 
\end{equation}
Assume by contradiction that there exists a sequence $s_l\to \infty$ such that
\begin{equation}\label{cal-C}
{\rm dist}_{H^1_{loc}}(\psi_b(s_l),{\bf S})\ge\delta,\quad\forall l\in\N
\end{equation}
for some $\delta>0$.  According to Corollary  \ref{coco},  there exist a subsequence $s_{l_n}$ of the sequence $s_l$, 
$\omega_1^+,\omega_2^+\in\R$ and vector-function $\gamma(x,t)$, defined in (\ref{gamma12}) such that the following convergence  hold
\[
\psi_b(x,t+s_{l_n})\to \gamma(x,t),\quad l_n\to\infty,\quad t\in\R.
\]
This implies that 
\begin{equation}\label{finn}
\psi_b(x,s_{l_n})\to\gamma(x,0)=\phi_{\omega_1^+}(x)+\phi_{\omega_2^+}(x),\quad l_n\to\infty,
\end{equation}
where 
$$
\phi_{\omega_1^+}(x)=C_1\!\left(\!\begin{array}{cc} e^{-\varkappa_1^{+}|x|}+\frac{me^{-\varkappa_1^{+}|x|}-\varkappa_1e^{-m|x|}}{\omega_1^+}\\ 
\varkappa_1^{+}\sgn x\frac{e^{-\varkappa_1^{+}|x|}-e^{-m|x|}}{\omega_1^+}\end{array}\!\right)\!,\quad
\phi_{\omega_2^+}(x)=C_2\!\left(\!\begin{array}{cc} -\varkappa_2^{+}\sgn x\frac{e^{-\varkappa_2^{+}|x|}-e^{-m|x|}}{\omega_2^+}\\
e^{-\varkappa_2^{+}|x|}-\frac{me^{-\varkappa_2^{+}|x|}-\varkappa_2^{+}e^{-m|x|}}{\omega_2^+} \end{array}\!\right).
$$
The convergence (\ref{finn}) contradict to (\ref{cal-C}).
This completes the proof of Theorem~\ref{main-theorem}.
\hfill$\Box$
\appendix
\section{Global well-posedness}
\label{sect-existence}
Here we prove Theorem~\ref{theorem-well-posedness}.
We first need to adjust the nonlinearity $F$ so that it becomes bounded, together with its derivatives. Define
\begin{equation}\label{def-Lambda}
\Lambda(\psi_0)=\sqrt{\frac{\mathcal{H}(\psi_0)-A}{m-B}},
\end{equation}
where $\psi\sb 0\in H^1$ is the initial data from Theorem~\ref{theorem-well-posedness} and ${A}$, ${B}$ are constants from (\ref{bound-below}).
Then we may pick a modified potential function $\widetilde{U}\in C\sp 2(\C^2)$, so that
\begin{equation}\label{new-U}
\widetilde{U}(\zeta)=U(\zeta)\qquad{\rm for}\ |\zeta|\le \Lambda(\psi_ 0),\quad\zeta\in\C^2,
\end{equation}
$\widetilde{U}(\zeta)$ satisfies (\ref{bound-below}) with the same constants ${A}$, ${B}$ as $U(\zeta)$ does:
\begin{equation}\label{new-U-2}
\widetilde{U}(\zeta)\ge {A}-{B}|\zeta|^2,\quad{\rm for}\ \zeta\in\C^2,\quad {\rm where}\ {A}\in\R\ {\rm and}\ 0\le {B}<m,
\end{equation}
and so that $|\widetilde{U}(\zeta)|$, $|\widetilde{U}'(\zeta)|$, and $|\widetilde{U}''(\zeta)|$ are bounded for $\zeta\in C^2$.
We define
$$
\widetilde{F}_j(\zeta)=-\partial_{\overline \zeta_j} \widetilde{U}(\zeta),\qquad\zeta\in\C^2,
$$
and  consider the Cauchy problem of type (\ref{KG-0}) with the modified nonlinearity,
\begin{equation}\label{KG-ap}
\left\{\begin{array}{l} 
i\dot\psi(x,t)=D_m\psi(x,t)-D_m^{-1}\delta(x)\widetilde F(\psi(0,t)),\qquad x\in\R, \quad t\in\R,
\\
\psi\at{t=0}=\psi_0(x),
\end{array}\right. 
\end{equation}
This is a Hamiltonian system, with the Hamilton functional
\begin{equation}\label{KG-a-h}
\widetilde{\mathcal{H}}(\psi)=\frac 12\langle\psi, (-\partial_x^2+m^2)\psi\rangle+\widetilde{U}(\psi(0,t)), \quad\psi\in H^1,
\end{equation}
which is Fr\'echet differentiable in the space $H^1$. 
By the Sobolev embedding theorem, $\Vert\psi\Vert_{L^\infty}^2 \le \frac 12\Vert\psi\Vert_{H^1}^2$. Moreover,
\begin{equation}\label{sobolev-embedding}
\Vert\psi\Vert_{L^\infty}^2 \le \frac1{2m}|\!\Vert\psi\Vert\!|^2,
\end{equation}
where $|\!\Vert\psi\Vert\!|^2:=\Vert\psi'\Vert_{L^2}^2+m^2\Vert\psi\Vert_{L^2}^2$.   Indeed, the Cauchy- Schwarz inequality and the Parseval identity imply
$$
\Vert\psi\Vert_{L^\infty}\le\frac 1{2\pi}\Vert\tilde\psi\Vert_{L^1} \le\frac 1{2\pi}\Big(\int |\tilde\psi(k)|^2(m^2+k^2) dk\Big)^{1/2}
\Big(\int \frac{dk}{m^2+k^2}\Big)^{1/2}
\le  \frac 1{2\pi}\sqrt{2\pi}|\!\Vert\psi\Vert\!|\sqrt{\frac {\pi}{m}}=\frac 1{\sqrt {2m}}\!\Vert\psi\Vert\!|.
$$
Thus (\ref{new-U-2}) leads to
$$
\widetilde{U}(\psi(0))\ge {A}-{B}\Vert \psi\Vert_{L^\infty}^2\ge {A}-\frac{B}{2m}|\!\Vert\psi\Vert\!|^2.
$$
Taking into account (\ref{KG-a-h}), we obtain the inequality
$$
|\!\Vert\psi\Vert\!|^2=2\widetilde{\mathcal{H}}(\psi)-2 \widetilde{U}(\psi(0))
\le 2\widetilde{\mathcal{H}}(\psi)-2{A}+\frac{B}{m}|\!\Vert\psi\Vert\!|^2,\qquad\psi\in H^1,
$$
which implies
\begin{equation}\label{t-bound-1}
|\!\Vert\psi\Vert\!|^2 \le\frac{2m}{m-B}\left(\widetilde{\mathcal{H}}(\psi)-{A}\right),\qquad\psi\in H^1.
\end{equation}
\begin{lemma}\label{lemma-same-u}
\begin{enumerate}
\item
$\widetilde{\mathcal{H}}(\psi_0)=\mathcal{H}(\psi_0)$.
\item
If $\psi\in H^1$ satisfies $\widetilde{\mathcal{H}}(\psi)\le\widetilde{\mathcal{H}}(\psi_0)$, then $\widetilde{U}(\psi(0))=U(\psi(0))$.
\end{enumerate}
\end{lemma}
\begin{proof}
\begin{enumerate}
\item
According to (\ref{sobolev-embedding}), (\ref{t-bound-1}), and the choice of $\Lambda(\psi_0)$ in (\ref{def-Lambda}),
$$
\Vert \psi_0\Vert_{L^\infty}^2\le \frac{1}{2m}|\!\Vert\psi_0\Vert\!|^2\le\frac{\mathcal{H}(\psi_0)-A}{m-B}=\Lambda^2(\psi_0).
$$
Thus, according to the choice of $\widetilde{U}$ (equality (\ref{new-U})),
$
\widetilde{U}(\psi\sb 0(0))=U(\psi\sb 0(0)),
$
proving ({\it i}).
\item
By (\ref{sobolev-embedding}), (\ref{t-bound-1}), the condition
$\widetilde{\mathcal{H}}(\psi)\le\widetilde{\mathcal{H}}(\psi_0)$, and part ({\it i}) of this lemma, we have:
\[
\Vert\psi\Vert_{L^\infty}^2 \le \frac{1}{2m}|\!\Vert\psi_0\Vert\!|^2\le\frac{{\mathcal H}(\psi)-A}{m-B}
\le\frac {{\mathcal H}(\psi_0)-A}{m-B}=\Lambda^2(\psi_0).
\]
Hence, ({\it ii}) follows by (\ref{new-U}).
\end{enumerate}
\end{proof}
\begin{remark}
We will show that if $\psi(t)$ solves (\ref{KG-ap}),
then $\widetilde{\mathcal{H}}(\psi(t))=\widetilde{\mathcal{H}}(\psi_0)$,
and therefore $\widetilde{U}(\psi(0,t))=U(\psi(0,t))$
by Lemma~\ref{lemma-same-u}~({\it ii}).
Hence, $\widetilde{F}(\psi(0,t))=F(\psi(0,t))$ for all $t\ge 0$,
allowing us to conclude that $\psi(t)$ solves (\ref{KG-0}) as well as (\ref{KG-ap}).
\end{remark}
\subsection*{Local well-posedness}
Denote by  $e^{-iD_m t}$  the dynamical group of the free Dirac equation. Then
the solution to the Cauchy problem (\ref{KG-ap}) can be represented by
\begin{equation}\label{integral-representation}
\psi(t)=e^{-iD_m t}\psi_0+Z[\psi(0,\cdot)](t),\quad
Z[\psi(0,\cdot)](t):=\int_0^te^{-iD_m(t-s)}D_m^{-1}\delta(\cdot)\widetilde{F}(\psi(0,s))\,ds,\quad t\in\R.
\end{equation}
The next lemma establishes the contraction principle for the integral equation (\ref{integral-representation}).
\begin{lemma}\label{lemma-bounds}
There exists a constant $C>0$ so that for any two functions $\psi_{k}(\cdot,t)\in C([-1,1],H^1)$, $k=1,\,2$, one has:
\[
\Vert{Z}[\psi_1(0,\cdot)](t)-{Z}[\psi_2(0,\cdot)](t)\Vert_{H^1}\le C|t|^{1/2} \sup_{|s|\le |t|}\Vert\psi_1(\cdot,s)-\psi_2(\cdot,s)\Vert_{H^1},\qquad |t|\le 1.
\]
\end{lemma}
\begin{proof}
It suffices to consider $t\ge 0$. In this case, 
\begin{equation}\label{FD}
e^{-iD_m t}=(i\partial_t+D_m){\cal G}(t),
\end{equation}
where ${\cal G}(t)$ is the integral operator with the integral kernel 
$$
{\cal G}(x,y,t)=G(x-y,t)=\theta(t-|x-y|)J_0(m\sqrt{t^2-(x-y)^2})/2.
$$ 
Here $J_0$ is the Bessel function.  According to  (\ref{integral-representation}) and (\ref{FD}),
\[
{Z}[\psi_1(0,\cdot)](t)-{Z}[\psi_2(0,\cdot)](t)=I_1(x,t)+I_2(x,t),
\]
where
\begin{eqnarray}\nonumber
I_1(x,t):&=&\int_0^t G(x,t-s)\left(\widetilde{F}(\psi_1(0,s))-\widetilde{F}(\psi_2(0,s))\right)\,ds,\\
\nonumber
I_2(x,t):&=&i\int_0^t [\dot G(\cdot,t-s)*D_m^{-1}\delta(\cdot)](x)(\widetilde{F}(\psi_1(0,s))-\widetilde{F}(\psi_2(0,s))\,ds 
\end{eqnarray}
First we prove the $L_2$ estimate for $I_j(x,t)$. By the Sobolev embedding theorem,
\begin{eqnarray}
\Vert I_1(\cdot,t)\Vert_{L^2}&\le& \sup_{s\in [0, t]} |\widetilde{F}(\psi_1(0,s))-\widetilde{F}(\psi_2(0,s))|\int_0^t\Vert G(\cdot,s)\Vert_{L^2}\,ds
\nonumber\\
\nonumber\\
&\le& C\sup_{z\in\C}|\nabla\widetilde{F}(z)|\Vert \psi_1(\cdot,s)-\psi_2(\cdot,s)\Vert_{H^1}\int_0^t \Vert \frac{\sin s\sqrt{\xi^2+m^2}}{\sqrt{\xi^2+m^2}}\Vert_{L^2}\,ds
\nonumber\\
\label{e-l2}
&\le& C_1\,t \sup_{s\in [0, t]}\Vert\psi_1(\cdot,s)-\psi_2(\cdot,s)\Vert_{H^1},
\end{eqnarray}
where we took into account that $|\nabla\widetilde{F}(z)|$ is bounded due to the choice of $\widetilde{U}$.
Similarly,
\begin{eqnarray}
\Vert I_2(\cdot,t)\Vert_{L^2}&\le&
C\sup_{s\in [0, t]}\Vert\psi_1(\cdot,s)-\psi_2(\cdot,s)\Vert_{H^1}\int_0^t\Vert (-i\xi \alpha+m\beta)\frac{\cos s\sqrt{\xi^2+m^2}}{\xi^2+m^2}\Vert_{L^2}\,ds
\nonumber\\
\label{e-l3}
&\le& C_1\,t \sup_{s\in [0, t]}\Vert\psi_1(\cdot,s)-\psi_2(\cdot,s)\Vert_{H^1}
\end{eqnarray}
Now, we derive the  $L^2$ estimates for the derivatives $\partial_x I_1(x,t)$ and $\partial_x I_2(x,t)$. We have
\[
\partial_x G(x,t)=\frac 1 2\theta(t-|x|)\partial_x J_0(m\sqrt{t^2-x^2})-\frac 12\delta(t-|x|)\sgn{x}.
\]
where  
$$
|\partial_x J_0(m\sqrt{t^2-x^2})|=|J_1(m\sqrt{t^2-x^2}) \frac{mx}{\sqrt{t^2-x^2}}|\le C,\quad  |x|\le |t|\le 1.
$$ 
Hence,
\begin{eqnarray}\label{d-e-l2}
\Vert\partial_x I_1(\cdot,t)\Vert_{L^2}&\le&
C_1\Vert\int_0^t (C\theta(t-s-|x|)+\delta(t-s-|x|))\,ds\Vert_{L^2}\,\sup_{s\in [0, t]} \Vert\psi_1(\cdot,s)-\psi_2(\cdot,s)\Vert_{H^1}
\nonumber\\
\nonumber\\
&\le& C_2[t\Vert\theta(t-|x|)\Vert_{L^2}+\Vert\theta(t-|x|)\Vert_{L^2} ]\,\sup_{s\in [0, t]} \Vert\psi_1(\cdot,s)-\psi_2(\cdot,s)\Vert_{H^1}
\nonumber\\
\nonumber\\
&\le& C_2 \,t^{1/2}(t+1)\,\sup_{s\in [0, t]}\Vert\psi_1(\cdot,s)-\psi_2(\cdot,s)\Vert_{H^1}.
\end{eqnarray}
Further,
$$
\partial_xD_m^{-1}\delta(x)=\partial_xD_mD_m^{-2}\delta(x)=
\frac{1}{2\pi} \int _{\R}\frac{e^{-i\xi x} (-i\xi)(-i\xi\alpha+m\beta)d\xi}{\xi^2+m^2}
=-\alpha\delta(x)+m^2 \alpha D_m^{-2}\delta(x)+m\beta \partial_x D_m^{-2}\delta(x)
$$
Hence,
\begin{eqnarray}\nonumber
\partial_x  I_2(x,t)&=& -i\alpha \int_0^t \dot G(x,t-s)\left(\widetilde{F}(\psi_1(0,s))-\widetilde{F}(\psi_2(0,s))\right)\,ds\\
\nonumber
&+&im^2\alpha\int_0^t [\dot G(\cdot,t-s)*D_m^{-2}\delta(\cdot)](x)\,\alpha(\widetilde{F}(\psi_1(0,s))-\widetilde{F}(\psi_2(0,s))\,ds\\
\nonumber
&+&im\beta\int_0^t \partial_x[\dot G(\cdot,t-s)*D_m^{-2}\delta(\cdot)](x)(\widetilde{F}(\psi_1(0,s))-\widetilde{F}(\psi_2(0,s))\,ds\\
\label{ek1}
&=& \alpha J_1(x,t)+\alpha J_2(x,t)+\beta J_3(x,t)
\end{eqnarray}
The $L^2$ norm of $J_1(x,t)$ is estimated similarly to  the $L^2$ norm of  $\partial_x I_1(x,t)$. Further, similarly to (\ref{e-l2}), we get
$$
\Vert J_2(\cdot,t)\Vert_{L^2}\le
C\sup_{s\in [0, t]}\Vert\psi_1(\cdot,s)-\psi_2(\cdot,s)\Vert_{H^1}\int_0^t\Vert \frac{\cos s\sqrt{\xi^2+m^2}}{\xi^2+m^2}\Vert_{L^2}\,ds
\le C_1\,t \sup_{s\in [0, t]}\Vert\psi_1(\cdot,s)-\psi_2(\cdot,s)\Vert_{H^1}
$$
$$
\Vert J_3(\cdot,t)\Vert_{L^2}\le
C\sup_{s\in [0, t]}\Vert\psi_1(\cdot,s)-\psi_2(\cdot,s)\Vert_{H^1}\int_0^t\Vert \frac{\xi \cos s\sqrt{\xi^2+m^2}}{\xi^2+m^2}\Vert_{L^2}\,ds
\le C_1\,t \sup_{s\in [0, t]}\Vert\psi_1(\cdot,s)-\psi_2(\cdot,s)\Vert_{H^1}
$$
\end{proof}
For $E>0$, let us denote $H^1_E=\{\psi_0\in H^1\sothat\mathcal{H}(\psi_0)\le E\}$.
\begin{corollary}
\label{cor-l8-existence}
\begin{enumerate}
\item
For any $E>0$ there exists $\tau=\tau(E)>0$ such that for any $\psi_0\in H^1_E$ there is a unique solution
$\psi(x,t)\in C([-\tau,\tau], H^1)$ to the Cauchy problem {\rm (\ref{KG-ap})} with the initial condition
 $\psi(0)=\psi_0$.
\item
The maps $W(t):\;\psi\sb 0\mapsto \psi(t)$, $t\in[-\tau,\tau]$ are continuous  from $H^1_E$ to $H^1$.
\end{enumerate}
\end{corollary}
\subsection*{Energy conservation and global well posedness}
\begin{lemma}\label{lemma-e-conserved}
For the solution to the Cauchy problem (\ref{KG-ap}) with the initial data
$\psi_0\in H^1$, the energy is conserved:
$\widetilde{\mathcal{H}}(\psi(t))=\const$, $t\in [-\tau,\tau]$.
\end{lemma}
\begin{proof}
The Galerkin approximations provide a solution $\psi\in L^\infty(\R,H^1)$ to (\ref{KG-ap})
with energy estimate
\begin{equation}\label{H-est}
\widetilde{\mathcal{H}}(\psi(t))\le  \widetilde{\mathcal{H}}(\psi_0),\qquad t\in\R.
\end{equation}
Moreover, estimates from the proof of Lemma \ref{lemma-bounds}   imply that 
 $\psi\in C(\R,H^1)$. Therefore,
$$
\mathcal{H}(\psi(t))=\mathcal{H}(\psi_0),\quad t\in [-\tau,\tau]
$$
since the inequality (\ref{H-est}) also holds with $\psi(s)$ instead of $\psi_0$ for every $s\in [-\tau,\tau]$
by the uniqueness of solutions proved in  Corollary \ref{cor-l8-existence}.
\end{proof}
\begin{corollary}
\label{coroll}
\begin{enumerate}
\item
The solution $\psi$ to the Cauchy problem (\ref{KG-ap}) with the initial data
$\psi\at{t=0}=\psi_0\in H^1$ exists globally: $\psi\in C_{b}(\R,H^1)$.
\item
The energy is conserved:
$\widetilde{\mathcal{H}}(\psi(t))=\widetilde{\mathcal{H}}(\psi_0),\qquad t\in \R$.
\end{enumerate}
\end{corollary}
\begin{proof}
Corollary~\ref{cor-l8-existence}~({\it i}) yields a solution
$\psi\in C([-\tau,\tau],H^1)$ with a positive $\tau=\tau(E)$.
However, the value of $\mathcal{H}(\psi(t))$ is conserved for $t\le \tau$
by Lemma~\ref{lemma-e-conserved}. Corollary~\ref{cor-l8-existence}~({\it i})
allows then to extend $\psi$ to the interval $[-2\tau,2 \tau]$, and eventually to all $t\in\R$.
\end{proof}
\subsection*{Conclusion of the proof of Theorem~\ref{theorem-well-posedness}}
The trajectory $\psi\in C_{b}(\R,H^1)$ is a solution to (\ref{KG-ap}), for which
Corollary~\ref{coroll}~({\it ii}) together with Lemma~\ref{lemma-same-u}~({\it i})
imply the energy conservation  (\ref{ec}).
By Lemma~\ref{lemma-same-u}~({\it ii}), $\widetilde{U}(\psi(0,t))=U(\psi(0,t))$, for all $t\in\R$.
This tells us that $\psi(x,t)$ is a solution to (\ref{KG-0}).
Finally, the a priori bound (\ref{eb}) follows from
(\ref{t-bound-1}) and the conservation of $\mathcal{H}(\psi(t))$.
This finishes the proof of Theorem~\ref{theorem-well-posedness}.
\section{Free Dirac equation}
\label{psi_pm_sol}
Here we show that the function
$$
\varphi^{\pm}(x,t)=\frac 1{2\pi}\int_\R \big(I+\frac{m}{\omega(k)}\beta\pm\frac{ik}{\omega(k)}\alpha\big) \frac{\hat f_d(\omega(k))}{2ik}
e^{\pm ikx}e^{-i\omega(k) t}\frac{kdk}{\omega(k)},\quad\omega(k)=\sqrt{k^2+m^2}
$$
are the solutions to the free Dirac equation (\ref{D-1}).
It suffices to prove that for any $q=(q_1,q_2)\in\C^2$ the  functions 
$$
p^{\pm}(x,t)=e^{\pm ikx}e^{-i\omega(k) t}\big(\omega(k)+m\beta\pm ik\alpha\big) q
$$
satisfy
$$
(i\partial_t-D_m)p^{\pm}(x,t)=0.
$$
Indeed, the functions
$u_j^{\pm}(x,t)=e^{\pm ikx}e^{-i\omega(k) t}q_j$ obviously satisfy
$$
\ddot u_j^{\pm}(x,t)+D_m^2 \,u_j^{\pm}(x,t)=0.
$$
Moreover,
$$
p^{\pm}(x,t)=(\omega(k)I+m\beta \pm ik\alpha) u^{\pm}(x,t)=(i\partial_t+D_m)u^{\pm}(x,t),\quad u^{\pm}(x,t)=(u_1^{\pm}(x,t),u_2^{\pm}(x,t)).
$$
Hence,
$$
(i\partial_t-D_m)p^{\pm}(x,t)=(i\partial_t-D_m)(i\partial_t+D_m)u^{\pm}(x,t)=-\ddot u^{\pm}(x,t)-D_m^2 \,u^{\pm}(x,t)=0.
$$
\section{Linear case}
\label{sect-linear-case}
Here we  consider the linear case, when 
\begin{equation}\label{F-lin}
F_j(\zeta_j)=a_j\zeta_j,\quad U_j(\zeta_j)=-\frac{a_j}2|\zeta_1|^2\qquad a_j\in\R.
\end{equation}
Now equation ({\ref{KG-0}) reads
\begin{equation}\label{KG-linear}
\dot\psi(x,t)=D_m\psi(x,t)-D_m^{-1}A\psi(0,t)\delta(x),\quad A=\left(\begin{array}{cc} a_1 & 0 \\ 0 & a_2 \end{array}\right),
\quad x\in\R, \quad t\in\R.
\end{equation} 
We restrict  our consideration to the case when $a_j<2m$, $j=1,2$. It is in this case that condition (\ref{bound-below}) is satisfied, and
then all conclusions of Theorem~\ref{theorem-well-posedness} on global well-posedness for equation (\ref{KG-linear}) hold .

Let us calculate  corresponding solitary waves. Now equations (\ref{qsol}) become
$$
\sqrt{m^2-\omega_1^2}=a_1\,\frac{\omega_1+m-\sqrt{m^2-\omega_1^2}}{2\omega_1},
\qquad \sqrt{m^2-\omega_2^2}=a_2\frac{\omega_2-m+\sqrt{m^2-\omega_2^2}}{2\omega_2},
\quad \omega_j\in (-m,m)
$$
Cancelling the nonzero factors $\sqrt{m+\omega_1}$ and $\sqrt{m-\omega_2}$, we obtain
$$
\sqrt{m-\omega_1}=a_1\,\frac{\sqrt{m+\omega_1}-\sqrt{m-\omega_1}}{2\omega_1},
\quad \sqrt{m+\omega_2}=a_2\,\frac{\sqrt{m+\omega_2}-\sqrt{m-\omega_2}}{2\omega_2}.
$$
Multiplying both sides  of this equations by $\sqrt{m+\omega_1}+\sqrt{m-\omega_1}$ and $\sqrt{m+\omega_2}+\sqrt{m-\omega_2}$, 
respectively, we get
\begin{equation}\label{aa}
\sqrt{m^2-\omega_1^2}+m-\omega_1=a_1,\qquad \sqrt{m^2-\omega_2^2}+m+\omega_2=a_2,\quad \omega_j\in (-m,m),
\end{equation}
where the left hand sides of both equalities are positive for $ \omega_j\in (-m,m)$.
Hence, there are no nonzero solitary waves for $a_j\le 0$, $j=1,2$.  
For $0<a_j<2m$, the corresponding  equation of  (\ref{aa}) has the  unique solution
\begin{equation}\label{w12}
\omega_j=\frac {(-1)^j}2 \big(a_j-m-\sqrt{m^2-a_j^2+2m a_j}\big).
\end{equation}
Finally,  we conclude that for $0<a_j<2m$, $j=1,2$, the set of finite energy solitary waves is given by
\begin{equation}\label{L}
{\cal S}=\left\{C_1\left[\begin{array}{c} e^{-\varkappa_1|x|}+\frac{me^{-\varkappa_1|x|}-\varkappa_1e^{-m|x|}}{\omega_1}\\\\
\varkappa_1\sgn x\frac{e^{-\varkappa_1|x|}-e^{-m|x|}}{\omega_1}\end{array}\right]
+C_2\left[\begin{array}{c} -\varkappa_2\sgn x\frac{e^{-\varkappa_2|x|}-e^{-m|x|}}{\omega_2}\\\\ 
e^{-\varkappa_2|x|}-\frac{me^{-\varkappa_2|x|}-\varkappa_2e^{-m|x|}}{\omega_2}\end{array}\right]
\sothat\; C_1,\,\,C_2\in\C \right\}.
\end{equation}
In the case $a_1<0$  the set ${\cal S}$ is given by (\ref{L}) with $C_1=0$, while $C_2\in\C$ is arbitrary, and vice versa.
.
\begin{theorem}\label{main-theorem-linear}
Assume that $F_j(\psi_j)=a_j\psi_j$, where $a_j<2m$, $j=1,2$. Then for any $\psi_0\in H^1$ the solution $\psi(t)\in C(\R,H^1)$
to the Cauchy problem (\ref{KG-0}) with $\psi(0)=\psi_0$ converges to ${\cal S}$ in the space $H^1_{loc}(\R)\otimes C^2$:
\begin{equation}\label{cal-A-linear}
\Psi(t)\to {\cal S}, \quad t\to \pm\infty.
\end{equation}
\end{theorem}
\begin{proof}
We proceed as in the proof of Theorem~\ref{main-theorem} until we get  equation (\ref{D-beta}), which takes now the following form:
\begin{equation}\label{D-beta-linear}
i\dot\gamma(x,t)=D_m\gamma(x,t)-D_m^{-1} A\gamma(0,t)\delta(x),\qquad (x,t)\in\R\sp 2.
\end{equation}
In this case  we cannot use  the Titchmarsh arguments since the condition (\ref{inv-u1}) fails.  Now  we should  prove that
\begin{equation}\label{fol}
\gamma(\cdot,t) \in {\cal S}, \qquad t\in\R,
\end{equation}
for all solutions of (\ref{D-beta-linear}) with the structure  (\ref{ber}).
In the Fourier transform $\hat\gamma(x,\omega)=\mathcal{F}\sb{t\to\omega}[\gamma(x,t)]$
the equation (\ref{D-beta-linear})  becomes
$$
(D_m-\omega)\hat\gamma(x,\omega)-D_m^{-1}A\hat\gamma(0,\omega)\delta(x)=0,\qquad (x,\omega)\in\R\sp 2.
$$
Applying the operator $D_m+\omega$, we get 
\begin{equation}\label{D-beta-linear-f}
(D^2_m-\omega^2)\hat\gamma(x,\omega)-A\hat\gamma(0,\omega)\delta(x)-\omega D_m^{-1}A\hat\gamma(0,\omega)\delta(x)=0,\qquad (x,\omega)\in\R\sp 2.
\end{equation}
On the other hand, the representation (\ref{ber}) implies that
$$
\hat\gamma(x,\omega)=\hat p(\omega)e^{-\varkappa(\omega)|x|}
+(\beta-\alpha \sgn x)\sigma^{-1}(\omega)\hat p(\omega)\frac{e^{-\varkappa(\omega)|x|}-e^{-m|x|}}{2\omega},\quad \hat\gamma(0,\omega)=\hat p(\omega).
$$
Substituting this into  (\ref{D-beta-linear-f}) and equating coefficients with delta functions, we obtain
$$
2\varkappa(\omega)\hat p(\omega)+\beta\sigma^{-1}(\omega)\hat p(\omega)\frac{\varkappa(\omega)-m}{\omega}=A\hat p(\omega),\quad
\varkappa=\sqrt{m^2-\omega^2},
$$
where $\sigma^{-1}(\omega)$ is the diagonal matrix (\ref{hatf})  with matrix elements 
$[\sigma^{-1}(\omega)]_{jj}:=\nu_{j}(\omega)=\varkappa(\omega)+m-(-1)^j\omega$.\\
Therefore, on the support of the distribution $\hat p_j(\omega)$, $j=1,2$, the identity hold
$$
2\varkappa(\omega)+(-1)^{j+1}\nu_{j}(\omega)\frac{\varkappa(\omega)-m}{\omega}=a_j.
$$
Simplifying, we arrive at the following equation
$$
\varkappa(\omega)+m+(-1)^{j}\omega=a_j.
$$
which are exactly   equations (\ref{aa}) for soliton parameters $\omega_j$.
Finally, we obtain that for $0<a_j<2m$
$$
\supp \hat p_j=\omega_j\in (-m,m),
$$
where $\omega_j$ are given in (\ref{w12}). Hence, the inclusion (\ref{fol}) follows.
This finishes the proof of Theorem~\ref{main-theorem-linear}.
\end{proof}

\end{document}